\documentclass[11pt]{article}

\usepackage{geometry}
\usepackage[numbers]{natbib}
\usepackage{amsfonts,amsmath,amssymb,amsthm}
\usepackage{mathrsfs}
\usepackage{lmodern}
\usepackage{graphicx,subcaption}
\usepackage[colorlinks,citecolor=blue,linkcolor=red,bookmarks=false,hypertexnames=true]{hyperref}
\usepackage[font=footnotesize,labelfont={bf,sf},labelsep=period]{caption}
\usepackage[table]{xcolor}
\usepackage{lettrine}
\usepackage[sf,bf]{titlesec}
\usepackage{enumitem}

\usepackage[bbgreekl]{mathbbol}
\usepackage{amsfonts}
\DeclareSymbolFontAlphabet{\mathbb}{AMSb}
\DeclareSymbolFontAlphabet{\mathbbl}{bbold}

\newtheorem{remark}{Remark}

\begin{document}

\title{Geodesic vortex detection on curved surfaces: Analyzing the 2002 austral stratospheric polar vortex warming event}

\author{F.\ Andrade-Canto\\ Departamento de Observaci\'on y Estudio de la Tierra, la Atm\'osfera y el Oc\'eano\\ El Colegio de la Frontera Sur\\ Chetumal, Quintana Roo, Mexico\\ \href{mailto:fernando.andrade@ecosur.mx}{\texttt{fernando.andrade@ecosur.mx}}\and F.J.\ Beron-Vera\\ Department of Atmospheric Sciences\\ Rosenstiel School of Marine, Atmospheric \& Earth Science\\ University of Miami\\ Miami, Florida, USA\\ \href{mailto:fberon@miami.edu}{\texttt{fberon@miami.edu}} \and G.\ Bonner\footnote{Now at Morgridge Institute for Research, Madison, Wisconsin, USA, \href{mailto:gbonner@morgridge.org}{\texttt{gbonner@morgridge.org}}.}\\ Department of Atmospheric Sciences\\ Rosenstiel School of Marine, Atmospheric \& Earth Science\\ University of Miami\\ Miami, Florida, USA}

\date{Started: June 14, 2024. This version: \today.\\ To appear in \emph{Chaos} as Featured Article.\vspace{-0.25in}}

\maketitle

\begin{abstract}
   Geodesic vortex detection is a tool in nonlinear dynamical systems to objectively identify transient vortices with flow-invariant boundaries that defy the typical deformation found in 2-d turbulence. Initially formulated for flows on the Euclidean plane with Cartesian coordinates, we have extended this technique to flows on 2-d Riemannian manifolds with arbitrary coordinates. This extension required the further formulation of the concept of objectivity on manifolds. Moreover, a recently proposed birth-and-death vortex framing algorithm, based on geodesic detection, has been adapted to address the limited temporal validity of 2-d motion in otherwise 3-dflows, like those found in the Earth's stratosphere. With these adaptations, we focused on the Lagrangian, i.e., kinematic, aspects of the austral stratospheric polar vortex during the exceptional sudden warming event of 2002, which resulted in the vortex splitting. This study involved applying geodesic vortex detection to isentropic winds from reanalysis data. We provide a detailed analysis of the vortex's life cycle, covering its birth, the splitting process, and its eventual death. In addition, we offer new kinematic insights into ozone depletion within the vortex.
\end{abstract}
 
\noindent\textbf{Geodesic vortex detection is a method for observer-independent (objective) identification from 2-d velocity data of vortices characterized by material (flow-invariant) boundaries that resist stretching over finite time, a notion formulated as a variational problem, hence the nomenclature. In this paper, geodesic vortex detection has been extended to operate on curved surfaces (manifolds), and a recently developed birth-and-death vortex framing algorithm derived from this method has been adapted to address the limited durational applicability of the 2-d motion assumption, a common practice in the Earth's stratosphere.  We used this modification to fully characterize the life cycle, from birth to death, of the austral stratospheric polar vortex during the sudden warming event that led to its splitting in 2002.  We also used it to provide novel kinematic perspectives into the mechanisms by which air within the vortex depletes ozone.  This analysis was carried out by examining wind velocity data from a reanalysis system (integrating past short-range forecasts with observations via data assimilation) on a constant potential temperature (i.e., isentropic) surface.} 

\section{Introduction}

The boundaries that deliniatecoherent Lagrangian vortices (CLVs) \cite{Haller-Beron-13, Haller-Beron-14} representLagrangian coherent structures (LCSs) \cite{Haller-Yuan-00, Haller-16, Haller-23} of the elliptic type. They extend the concept of invariant tori in time-periodic or quasi-periodic 2-d flows \cite{Ottino-89} to time-aperiodic 2-d flows. The flow-invariant (i.e., material) boundaries of CLVs have the characteristic that each of itssubsetsstretches equally over a finite-time interval $[t_0,t_0+T]$. This property allows them to resist the exponential stretching that arbitrary material loops typically experience in 2-d turbulence. The methodology for identifying CLVs is known as \emph{geodesic vortex detection}, following the terminology introduced in \cite{Beron-etal-12}. These authors adopted the variational formulation developed in \cite{Haller-Beron-12}, which evolved into the one in \cite{Haller-Beron-13}. Roughly speaking, the relevant action represents an averaged measure of relative stretching, and the material loops that form the boundaries of CLVs are solution curves, i.e., geodesics, that extremize this action. This further enabled an analogy between CLVs and black holes in cosmology, as elaborated in \cite{Haller-Beron-14}.  In the variational theories of LCSs, material deformation is quantified using the Cauchy--Green strain tensor, known in fluid mechanics as an objective, or material frame-indifferent, tensor \cite{Speziale-98}. The objectivity of this tensor is integral to geodesic vortex detection, allowing long-lasting vortices to be identified independently of the observer viewpoint. Geodesic vortex detection has been widely applied, predominantly in oceanographic settings \cite{Wang-etal-15, Wang-etal-16, Beron-etal-15, Andrade-etal-20, Andrade-etal-22, Andrade-Beron-22, Bodnariuk-etal-24}, with exceptions in Earth \cite{Serra-etal-17} and planetary \cite{Hadjighasem-Haller-16} atmospheric applications.

Geodesic vortex detection, as extended in \cite{Andrade-etal-20}, allows identifying the birth and death of CLVs without predefined parameters. This is achieved by thoroughly exploring the space $(t_0, T)$, where $t_0$ rolls over a sufficiently wide time window to cover the vortex's lifetime that we aim to detect. For each $t_0$, $T$ is extended as long as geodesic vortex detection succeeds. In this way, for each $t_0$, we determine a life expectancy $T_\text{exp}(t_0)$, which represents the maximum $T$ for which a vortex is geodesically detected at $t_0$. Ideally, a linearly decaying (wedge-shaped) $T_\text{exp}(t_0)$ emerges, suggesting that all Lagrangian coherence assessments consistently predict the vortex's death. To detect the birth of a CLV, the aforementioned procedure must be executed in reverse chronological order.

The direct implementation of geodesic vortex detection and the birth-and-death CLV framing algorithm on Earth's stratospheric flows, which prompted this research, presents complexity due to two primary factors. First, the curvature of the Earth plays a significant role in stratospheric flows on isentropic (constant-potential-temperature) hemispherical caps centered at the poles. However, geodesic vortex detection was originally derived for flows on a Euclidean plane with Cartesian coordinates, necessitating its adaptation. Second, the assumption of 2-d flows holds limited validity in the stratosphere. It has been observed \cite{Haynes-05} that after an approximate duration of one month, air parcels initially located on a specific isentropic surface begin to experience notable diapycnic (cross-isentropic) mixing. However, the birth-and-death CLV framing does not prescribe a specific duration $T$ for assessing coherence, thus requiring modification.

\subsection{Stratospheric polar vortices, ozone holes, and sudden warmings}

\begin{table}[]
    \centering
    \begin{tabular}{ll}
         \hline\hline
         CLV & Coherent Lagrangian Vortex \\
         LCS & Lagrangian Coherent Structure\\
         SPV & Stratospheric Polar Vortex\\
         SSW & Sudden Stratospheric Warming\\
         \hline
    \end{tabular}
    \caption{List of most frequently used acronyms in the paper.}
    \label{tab:acro}
\end{table}

The stratosphere features a layered temperature structure with rising temperatures from ultraviolet radiation absorbed by the ozone layer \mbox{\cite{Andrews-etal-87}}. It begins around 20 km at the equator, 10 km at mid-latitudes, and 7 km at the poles.  Winds can reach speeds of up to 60 m\,s$^{-1}$ along the polar night jet near 50$^\circ$ latitude around the stratospheric polar vortex (SPV) \mbox{\cite{Waugh-Polvani-10}}.  (Refer to Table \mbox{\ref{tab:acro}} for a list of most frequently employed acronyms in this manuscript.)\\\indent The SPV forms in autumn as temperatures decrease rapidly with the onset of the polar night. The resulting temperature gradient between the poles and the tropics, mediated by the Coriolis force, causes the vortex to rotate cyclonically. The austral SPV is more resistant to disruptions from upward-propagating Rossby waves, which are less impactful in the austral atmosphere due to fewer topographic features from its smaller landmass. The vortex confines \mbox{\cite{Juckes-McIntyre-87}} chemical reactions on polar stratospheric clouds, activating chlorine and causing ozone depletion in late spring when temperatures are low but light is sufficient, producing an ozone hole \mbox{\cite{Solomon-99}}. Chlorofluorocarbons (or CFCs) catalyze the destruction of ozone \mbox{\cite{Molina-Rowland-74, Farman-etal-85}}.\\\indent Sudden stratospheric warmings (SSWs) \mbox{\cite{Matsuno-71}} can significantly weaken the SPV \mbox{\cite{Shepherd-00, Polvani-Waugh-04, Haynes-05}}. They are minor if the meridional temperature gradient reverses, and major if the polar night jet also reverses \mbox{\cite{Butler-etal-15}}. Major SSWs mainly occur in the Northern Hemisphere, about six times per decade \mbox{\cite{Jucker-etal-21}}. Only one has been recorded in the Southern Hemisphere in 2002, causing the SPV to split \mbox{\cite{Shepherd-etal-05}}, with a weaker SSW later reported \mbox{\cite{Lim-etal-21}}.This paper concentrates on the \emph{kinematic} aspects of the austral SPV as affected by the SSW in 2002, deliberately omitting the significant but distinct fluid dynamics explanations \cite{Esler-etal-06}. We examine the kinematics through the lens of nonlinear dynamical systems.

\subsubsection{Kinematics of SVPs}\label{sec:spvs}

The boundary, more commonly referred to as ``edge'' \cite{Nash-etal-96}, of the austral SVP was classified as a shearless LCS \cite{Farazmand-etal-14} in \cite{Rypina-etal-07a, Beron-etal-08-JAS}, heuristically using finite-time Lyapunov exponents (FTLEs) \cite{Haller-Yuan-00}. Beyond the heuristics of the FTLE method, the work of \cite{Rypina-etal-07a, Beron-etal-08-JAS} explained the resilience to breakup of the SPV edge, that is, its role as a barrier for the transport of chemical species,  using concepts from Kolmogorov--Arnold--Moser (KAM) theory \cite{Arnold-etal-06}, extended to scenarios in which the twist condition is violated, notably occurring at the core of the polar night jet, and where the nonintegrable Hamiltonian perturbation is multiply-periodic in time \cite{Rypina-etal-07b}. The tendency of these degenerate KAM tori to resist disintegration was previously examined by \cite{delCastillo-Morrison-93} in the context of time-periodicity. Although kinematic in essence, the explanation provided by KAM theory was shown to be superior to the inherently dynamical explanation based on the potential vorticity (PV) argument advocated by \cite{McIntyre-89, Dritschel-McIntyre-08} regarding the elasticity and resilience of PV isolines, as it remains applicable even for easterly stratospheric zonal jets \cite{Beron-etal-12}, where PV is homogeneous throughout.

More recently, \cite{Serra-etal-17} studied the kinematics of the boreal SPV during the exceptionally cold winter of 2013--2014 in northeastern North America. In that study, the edge of the SPV was rigorously classified as an elliptic LCS through the application of geodesic vortex detection. The elliptic LCSs, computed at different isentropic surfaces and times, were situated always equatorward of the polar night jet, thereby providing a more precise delineation of the SPV edge than shearless LCSs. The significance of the unveiled Lagrangian edge was corroborated by the notable contrasts in temperature and ozone concentration across it. The authors also illustrated that PV-based methods \cite{Nash-etal-96} incorrectly delineate the true extent of the vortex.  However, \cite{Serra-etal-17} applied geodesic vortex detection without explicitly addressing how they accounted for the Earth's sphericity.

\subsubsection{Kinematics of SSWs}

Using a different tool from nonlinear dynamical systems, other than FTLE and geodesic vortex detection, the SSWs that precipitated the splitting of the austral stratospheric SPV in 2002 \cite{Curbelo-etal-19} and the boreal SPV in 2020 \cite{Curbelo-etal-21} were examined. The adopted tool, referred to as the $M$-function \cite{Mancho-etal-13}, was used by the authors to propose a definition for the SPV's edge through structures akin to hyperbolic LCSs \cite{Haller-Beron-12, Haller-16}. Hyperbolic LCSs are devoid of shear \cite{Farazmand-etal-14}; however, contrary to shearless LCSs, they normally attract or repel proximate fluid trajectories in forward time when computed in reverse, and vice versa when considered in forward time. As a result, they undergo stretching or contraction distinctively more prominently than shearless and elliptic LCSs, which do so minimally, serving as finite-time analogs of invariant tori, which naturally delineate the boundaries of Lagrangian vortices. Using hyperbolic LCSs to define the SPV's edge is an unconventional approach motivated by the analysis of autonomous dynamical systems, wherein stable and unstable manifolds of saddle points can form heteroclinic trajectories enclosing centers \mbox{\cite{Ottino-89}}. On the other hand, the $M$-method, relying on trajectory length, lacks objectivity and, similar to the FTLE method, can only approximate hyperbolic LCSs heuristically. The FTLE method does this by identifying ridges in the FTLE field, which can find rigorous support under certain conditions \mbox{\cite{Haller-11}}. The $M$-function method identifies such structures through arbitrary thresholding of the ``gradient'' of the $M$-function.Despite these limitations, \cite{Curbelo-etal-21} were able to describe distinctive aspects of the bifurcation of the boreal SPV, such as the aggregation of ozone-depleted air predominantly within the primary of the resultant vortices of the SSW.

It is important to acknowledge that \cite{Lekien-Ross-10} conducted an FTLE analysis of the austral SPV in 2002 during the SSW, primarily as a proof of concept for the computation of FTLE fields on Riemannian manifolds. Furthermore, \cite{Atnip-etal-24} examined the bifurcation of the austral SPV due to the SSW in the same year, aiming to demonstrate the concept of semi-material finite-time coherent sets as introduced in \cite{Froyland-Koltai-23}, which is derived from the spectral analysis of an ``inflated'' dynamic Laplacian where time is allowed to become a diffusion process itself.  Within the probabilistic transfer-operator-based approach in a set-oriented numerical framework, the work of \cite{Ndour-etal-21} on early warnings of sudden changes in flow patterns, like the SPV splitting in 2002, is also noteworthy.

\subsection{This paper}

The aim of this paper is twofold.  First, we extend geodesic vortex detection to flows on 2-d Riemannian manifolds using arbitrary coordinates. Distinguishing between coordinate and metric representations of vectors and tensors is crucial and has been highlighted. This extension led us to further develop the concept of objectivity on manifolds. The birth-and-death framing algorithm based on geodesic vortex detection is modified to handle the temporary validity of 2-d motion in mainly 3-d flows, like those in the Earth's stratosphere. 

We then analyze the kinematics of the austral SPV during the intense 2002 SSW event, known for vortex splitting, using geodesic vortex detection on isentropic winds from reanalysis data. This includes a detailed analysis of the vortex’s life cycle---its birth, splitting, and death---and provides new kinematic insights into ozone depletion within the vortex.  As part of this contribution, we provide a code to the community, developed based on publicly available software, incorporating the necessary adaptations for application of geodesic vortex detection to flows on manifolds.

The structure of this paper is as follows. Section \ref{sec:M} explores an extended approach to geodesic vortex detection, suitable for identifying CLVs within flows on 2-d manifolds, along with an adaptation of the objectivity concept and modifications to the algorithm to track the birth and death of CLVs relevant to the transient validity of the 2-d flow model. Section \ref{sec:life} provides a comprehensive analysis of the SPV's life cycle in 2002. In Section \ref{sec:o3}, we examine the transport barrier characteristics of the edge of the SPV, as identified geodesically, through ozone concentration, focusing on the kinematics of depletion within the SPV. The paper concludes in Section \ref{sec:conclusion} with a summary and final observations. In addition, two appendices are included, offering mathematical (Appendix \ref{app:loops}) and computational (Appendix \ref{app:num}) information.

\section{Geodesic vortex detection on curved surfaces}\label{sec:M}

We start by extending geodesic vortex detection to 2-d manifolds. Although related work has been discussed in the literature \cite{Lekien-Ross-10, Beron-etal-10b, Karrasch-15}, it does not address the identification of CLVs.

\subsection{Preparation}

It is presumed that the reader has a basic understanding of differential geometry; for underlying details not explicitly covered here, reference may be made to \cite{Abraham-Marsden-Ratiu-88}. Let $M$ be a smooth two-dimensional manifold with a smooth embedding $\iota:M\hookrightarrow\mathbb R^3$. Given a local chart $\chi:U\subset M\to V\subset\mathbb R^2$, a point $p\in U$ has coordinates $x=\chi(p)=(x^1,x^2)$. The corresponding local parametrization is
\begin{equation}
    \varphi:=\iota\circ\chi^{-1}:V\to\mathbb R^3,
    \label{eq:phi}
\end{equation}
which we write as
\begin{equation}
    \varphi(x)=
    \begin{pmatrix}
        \varphi^1(x)\\
        \varphi^2(x)\\
        \varphi^3(x)
    \end{pmatrix}.
\end{equation}
Hereafter, we identify a point $p\in U$ with its coordinate pair $x=\chi(p)$ and use $x$ to denote either.

Consider a vector $v$ on $M$ at $x$.  More precisely, consider $v\in T_xM$ where $T_xM$ is the tangent space to $M$ at point $x$.  We write $v$ in boldface, $\mathbf v$, to denote its \emph{coordinate representation}, that is, the 2-tuple formed by the components of $v$ in the (global) coordinates $x$ on $M$. Let $g(x) : T_xM\times T_xM \to \mathbb R$ be a \emph{Riemannian metric} on $M$, i.e., a positive-definite symmetric bilinear form on $M$.  One can obtain it via \emph{pullback} by $\varphi$ of the Euclidean metric ($e$) of the ambient space, namely, $e(\varphi(x)) = (\varphi^*e)(x) = g(x)$.  Explicitly, for any (tangent) vectors $v,w\in T_xM$ one has
\begin{equation}
    g(x)(v,w) = e(\varphi(x))(D\varphi(x)\cdot v,D\varphi(x)\cdot w) = \mathbf v\cdot G(x)\mathbf w,
    \label{eq:g}
\end{equation}
where the differential $D\varphi(x)$ is the linear map $T_xM \to T_{\varphi(x)}M$, whose coordinate representation is the 3-by-2 matrix $\nabla\varphi(x)$, and the 2-by-2 matrix
\begin{equation}
    G(x) := \nabla\varphi(x)^\top\nabla\varphi(x)
\end{equation}
is the coordinate representation of $g(x)$. In \eqref{eq:g}, the symbol $\cdot$ is used to represent both the application of a linear map and the Euclidean inner product, which is a common practice.  Positive-definiteness of $G(x)$, and hence of $g(x)$, follows from invertibility of $\varphi$. 

We use blackboard bold letters to denote the \emph{metric representation} of vectors. For instance, for $v\in T_xM$, 
\begin{equation}
    \mathbbl v(x) = G(x)^{1/2}\mathbf v.
    \label{eq:phys}
\end{equation}
The square root of $G$ (temporarily omitting the dependence on $x$ for simplicity) may be interpreted as follows. Given that $G$ is symmetric and positive-definite, its eigenvalues are positive and its eigenvectors are mutually orthogonal. Consequently, $G$ can be decomposed into $G = VDV^\top$, where $D$ is the diagonal matrix whose diagonal entries are the eigenvalues of $G$, and $V$ is the matrix whose columns consist of the corresponding eigenvectors of $G$. The latter satisfies $V^\top V = \mathsf I = VV^\top$, where $ \mathsf I$ denotes the identity matrix. Under these conditions, it follows that $G^{1/2} = VD^{1/2}V^\top$, since $(VD^{1/2}V^\top)^2 = G$, as anticipated. Furthermore, the following holds true: 
\begin{equation}
    G(x)^{-1/2}G(x)G(x)^{-1/2} = \mathsf I,
    \label{eq:phys-G}
\end{equation} 
where $G^{-1/2} = (VD^{1/2}V^\top)^{-1} = VD^{-1/2}V^\top$. The magnitude of $v\in T_xM$ is thus calculated as 
\begin{equation}
    \|v\|_x = \sqrt{\mathbf v \cdot G(x)\mathbf v} = \sqrt{\mathbbl v(x)\cdot \mathbbl v(x)}.
\end{equation}
This result relies on \eqref{eq:phys-G}, which asserts that $\mathsf I$ serves as the metric representation of $g(x)$. It is noteworthy, albeit somewhat redundant, that in metric coordinates, the computation of the inner product is done using the Euclidean inner product.

\begin{remark}
    When $M = \mathbb R^2$, $G = \mathsf I$ and the distinction between coordinate and metric representations is immaterial.  In that case, for instance, $\mathbbl v = \mathbf v$.
\end{remark}

\subsection{Coherent Lagrangian vortices on $M$}

Consider a fluid contained in an open domain $\mathcal D\subset M$. Let $u(x,t)$ be its time-dependent velocity field. We will assume that $t \in \mathcal I := [t_-,t_+] \subset \mathbb R$, as is the case when dealing with observational data.  Let $F^{t_0,t_1}(x_0)$ denote the solution of the fluid particle motion equation
\begin{equation}
  \dot x = u(x,t)
  \label{eq:dxdt}
\end{equation}
for initial condition $x(t_0) = x_0$.  That is,
\begin{equation}
    F^{t_0,t} : \mathcal D \to \mathcal D
    \label{eq:F}
\end{equation}
is the flow map that relates time-$t_0$ and $t$ positions of fluid particles, satisfying the group law $F^{t_0,t_0} = \text{id}$ and $F^{t_0,t} = F^{t',t}\circ F^{t_0,t'}$ where $t_0 < t' < t \in \mathcal I$. We assume that $u(x,t)$ is sufficiently smooth in each argument to ensure the existence of $F^{t_0,t}$ and its inverse, $(F^{t_0,t})^{-1} = F^{t,t_0}$. Established results from differential equation theory \cite{Arnold-73} then guarantee that $F^{t_0,t}(x_0)$ is as smooth in $x_0$ as $u(x,t)$ is in $x$.

The notion of a vortex with a material, i.e., flow-invariant, boundary resisting stretching under advection from time $t_0 \in \mathcal I$ to time $t_1 := t_0 + T \in \mathcal I$ for some (finite) $T$ is expressed by a variational principle \cite{Haller-Beron-13, Haller-Beron-14}.  Specifically, let $\gamma$ be a material loop at time $t_0$ and $F^{t_0,t_1}(\gamma)$ be its image at time $t_1$ under advection.  Let $S^1 \simeq \gamma \ni s \mapsto r(s) \in \mathcal D$ be a parameterization for $\gamma \hookrightarrow \mathcal D$. The pointwise relative stretching experienced by the loop when evolving from $t_0$ to $t_1$ is given by
\begin{equation}
    L(r, r') := \frac{\|dF^{t_0,t_1}(r)\|_{F^{t_0,t_1}(r)}}{\|dr\|_{r}} = \frac{\sqrt{\mathbf r'\cdot  C^{t_0,t_1}(r) \mathbf r'}}{\sqrt{\mathbf r'\cdot G(r) \mathbf r'}},
    \label{eq:L}
\end{equation}
where $r'(s) = \frac{dr}{ds} \in T_{r(s)}\mathcal D$ with $\mathbf r'(s)$ being its coordinate representation, following the notation introduced above.  Moreover,  with the indices $t_0,t_1$ dropped for notational simplicity,
\begin{equation}
    C(x) := \nabla F(x)^\top G\big(F(x)\big)\nabla F(x)
    \label{eq:C-coordinate}
\end{equation}
is the coordinate representation of the (right) Cauchy--Green strain tensor, $c(x)$. Due to the invertibility of the coordinate map \eqref{eq:phi} and the flow map $F^{t_0,t}$, it follows that $C(x)$, and consequently $c(x)$, is positive-definite. Geometrically, $c(x)$ represents a Riemannian metric induced by $F$ via pullback of $g(x)$.\footnote{This interpretation was first discussed in \cite{Haller-Beron-12}, but in the $M = \mathbb R^2$ case. In Appendix G of \cite{Haller-Beron-12}, $c$ is interpreted as the pullback by $\varphi \circ F$ of $e$. However, there is a misinterpretation regarding the view of $\varphi$ as a map $\mathbb{R}^2 \to M$. A correct interpretation of $c$ as the pullback by $F$ of $g$ is given in \cite{Karrasch-15} in the context of hyperbolic LCSs. More recent interpretations appear in \cite{Froyland-Koltai-23, Atnip-etal-24}, but these interpretations suggest a focus shift away from $c$ in material coherence detection.} That is, $g(F(x)) = (F^*g)(x) = c(x)$, hence
\begin{equation}
    c(x)(u,v) = g(F(x))(DF(x)\cdot v,DF(x)\cdot w) = \mathbf v\cdot C(x)\mathbf w = \mathbbl v(x)
    \cdot \mathsf C(x)\mathbbl w(x)
\end{equation}
for any $v,w\in T_x\mathcal D$, where
\begin{equation}
    \mathsf C(x) := G(x)^{-1/2}C(x)G(x)^{-1/2}
    \label{eq:C-metric}
\end{equation}
gives the metric representation of $c(x)$. Observe the difference in the way the argument of $G$ is evaluated: in \eqref{eq:C-metric}, it is $G(x)$, while in \eqref{eq:C-coordinate}, it is $G(F(x))$. Additionally, using \eqref{eq:C-metric}, \eqref{eq:L} can be equivalently expressed as
\begin{align}
    L(r,r') = \frac{\sqrt{\mathbbl r'\cdot  \mathsf C^{t_0,t_1}(r) \mathbbl r'}}{\sqrt{\mathbbl r'\cdot \mathbbl r'}}.
\end{align}

The variational principle proposed by \cite{Haller-Beron-13, Haller-Beron-14} is formulated as
\begin{equation}
  \mathfrak S_L[r] := \oint_\gamma L(r(s), r'(s))\,ds,\quad 
  \left.\frac{d}{d\epsilon}\right\vert_{\epsilon = 0} \mathfrak S_L[r + \epsilon n] = 0,
  \label{eq:variational}
\end{equation}
where $n(s) \in T_{r(s)}\mathcal D$ is normal to $\gamma$.  The action $\mathfrak S_L[r]$ in \eqref{eq:variational} measures relative stretching from $t_0$ to $t_1$ on average along $\gamma$.  The corresponding Lagrangian, $L(r,r')$, is invariant under $s$-shifts and thus represents a Noether quantity.  That is, it is equal to a positive constant, say $p$.  This means that solutions to the variational principle \eqref{eq:variational} are characterized by uniformly $p$-stretching loops. Embedded within $O(\varepsilon)$-thick coherent material belts showing no observable variability in averaged relative stretching, the time-$t_0$ positions of such \emph{$p$-loops} turn out to be limit cycle solutions to
\begin{equation}
  r' = 
  \ell_p^\pm(r) :=
  \sqrt{
  \frac
  {\lambda_2(r) - p^2}
  {\lambda_2(r) - \lambda_1(r)}
  }
  \,\nu_{(1)}(r) 
  \pm
  \sqrt{
  \frac
  {p^2 - \lambda_1(r)}
  {\lambda_2(r) - \lambda_1(r)}
  }
  \,\nu_{(2)}(r),  
  \label{eq:p-line}
\end{equation}
for either $\ell^p_+(r)$ or $\ell^p_-(r)$, which represent bidirectional vector or \emph{line} fields.

In \eqref{eq:p-line}, $\lambda_1(r) < p^2 < \lambda_2(r)$ and $\{\nu_{(i)}(r)\}$ satisfy (ignoring dependencies on $r$ for brevity)
\begin{equation}
    G^{-1}C\boldsymbol\nu_{(i)} = \lambda_i\boldsymbol\nu_{(i)} \Longleftrightarrow \mathsf C\bbnu_{(i)} = \lambda_i\bbnu_{(i)},\quad i=1,2.
    \label{eq:evec}
\end{equation}
Note that $(\lambda_i,\bbnu_{(i)})$ is the $i$th eigenvalue--eigenvector pair of the metric representation of the Cauchy--Green stress tensor. However, $(\lambda_i, \boldsymbol{\nu}_{(i)})$ is \emph{not} that of its coordinate representation, $C$; rather, it is of the matrix $G^{-1}C$. Due to the symmetry and positive definiteness of $C$ and $G$, it follows that $0 < \lambda_1 \le \lambda_2$.  Moreover, 
\begin{equation}
    \boldsymbol\nu_{(i)}\cdot G\boldsymbol\nu_{(j)} = \bbnu_{(i)}\cdot \bbnu_{(j)} = \delta_{ij},\quad i,j = 1,2,
    \label{eq:ortho}
\end{equation}
which holds under an appropriate normalization.  That is, the (orientationless) eigenvectors of $G^{-1}C$ and $\mathsf C$ are orthogonal, as measured in the coordinate and metric spaces, respectively.

\begin{remark}
    The eigenvalue--eigenvector pairs $(\bar\lambda_i,\bar{\boldsymbol\nu}_{(i)})$, $i=1,2$, of $C$ differ from those of $G^{-1}C$ and $\mathsf C$. While the eigenvalues satisfy $0 < \bar\lambda_1 \le \bar\lambda_2$, and the eigenvectors $\bar{\boldsymbol\nu}_{(i)}\cdot \bar{\boldsymbol\nu}_{(j)} = \delta_{ij}$, the latter cannot be interpreted as an orthogonality condition on $M$, unless $M = \mathbb{R}^2$, in which case all these differences become irrelevant.
\end{remark}

To confirm that $r' = \ell_p^\pm(r)$ satisfies $L(r, r') = p$, one can refer to \cite{Haller-Beron-13}, while considering the orthonormality relationship \eqref{eq:ortho}. Limit cycle solutions to equation \eqref{eq:p-line}, or $p$-loops, form smooth annular regions of nonintersecting loops. The outermost loop in a set of $p$-loops represents a material closed curve which delineates the boundary of a \emph{coherent Lagrangian vortex} (or \emph{CLV}).

Appendix \ref{app:loops} provides additional details about $p$-loops. The algorithmic procedures related to geodesic vortex detection, including the adaptations presented here, are detailed in Appendix \ref{app:num}.

\begin{remark}[On the origin of the geodesic vortex detection terminology]
    The $p$-loops can also be interpreted as so-called null-\emph{geodesics} of the (sign-indefinite) generalized Green--Lagrangian tensor field, with coordinate and metric representations, respectively, given by
    \begin{equation}
        C_p(x) := C(x) - pG(x)  \Longleftrightarrow \mathsf C_p(x) := \mathsf C(x) - p\mathsf I.  
    \end{equation}
    This follows by applying the same steps as in supplementary Appendix B of \textup{\cite{Haller-Beron-13}}, which leads to the conclusion that stationary curves of $\mathfrak S_L[r]$ are also of $\mathfrak S_{E_p}[r]$ where $E_p(r, r') := \mathbf r'\cdot C_p(r) \mathbf r' = \mathbbl r'\cdot \mathsf C_p(r) \mathbbl r' \equiv 0$. This and related results \textup{\cite{Haller-Beron-12, Beron-etal-13, Farazmand-etal-14, Haller-15, Haller-23}} explain the \textbf{geodesic vortex detection} terminology.
\end{remark}

\subsection{Objectivity of geodesic vortex detection on $M$}

The Lagrangian \eqref{eq:L} measures pointwise relative stretching in an observer-independent manner or, in fluid mechanics jargon, \emph{objectively} on $M$.  To see this, one needs to verify that it remains the same under the general observer change of stand point from $x$ to $\bar x$, defined by
\begin{equation}
    \bar x = \Phi(x,t) =: \Phi^t(x),
    \label{eq:Q}
\end{equation}
where $\Phi^t : M \to M$ is some smoothly invertible map. Let
\begin{equation}
    Q^t(x) := \nabla \Phi^t(x),
\end{equation} 
which is the coordinate representation of the differential $D\Phi^t(x)$, which can be viewed as a linear map $T_xM \to T_{\bar x}M$. Dropping the $t$-index from $Q^t$ herein for brevity, this satisfies
\begin{equation}
    Q(x)^\top\bar G(\bar x) Q(x) = G(x). 
    \label{eq:Q-rot}
\end{equation}
This describes a change of observers that aligns with the metric $g(x)$ on $M$.  

To see the above, recall that a vector on $M$ is said to be objective if different observers see it the same way albeit perhaps oriented differently \cite[][Chapter 3]{Haller-23}.  This condition is fulfilled for a \emph{constant} vector $v\in T_xM$ because it is mapped by $\Phi^t$ to $\bar v\in T_{\bar x}M$ via pushforward as
\begin{equation}
    \bar v = \Phi^t_*v = D\Phi^t\cdot v\Longleftrightarrow  \bar{\mathbf v} = Q \mathbf v,
    \label{eq:obj}
\end{equation}
where dependencies on $x$ and $\bar x$ have been dropped for simplicity.  With this in mind, we compute
\begin{equation}
    \|\bar v\|_{\bar x}^2 = \bar{\mathbf v}\cdot \bar G\bar{\mathbf v} = \mathbf v\cdot Q^\top \bar G Q \mathbf v = \mathbf v\cdot G\mathbf v = \mathbbl v\cdot \mathbbl v = \|v\|_x^2
\end{equation} 
upon using \eqref{eq:phys-G} and \eqref{eq:Q-rot}.  Thus a constant vector on $M$ is objective as it had been anticipated. The equation on the right-hand side of \eqref{eq:obj} is the required condition for the coordinate representation of \emph{any} vector on $M$ to ensure the vector's objectivity. It is noteworthy that not all vectors on $M$ fulfill this condition. A important example is the fluid velocity $u(x,t) =: u_t(x) = (\partial_tF^{t_0,t}\circ F^{t,t_0})(x)$, which transforms under \eqref{eq:Q} as $\bar u_t = \partial_t\Phi^t + D\Phi^t\cdot u_t$.
 
Now, returning to show that \eqref{eq:L} objectively measures pointwise relative stretching on $M$, first note that
\begin{equation}
    d\bar r = D\Phi^t\cdot dr\Longleftrightarrow d\bar{\mathbf r} = Q d\mathbf r,
\end{equation}
which demonstrates that $r'(s) \in T_{r(s)}\mathcal D$ is objective, and the denominator in equation \eqref{eq:L} is measured uniformly by both the $x$- and $\bar{x}$-observers: 
\begin{equation}
    \|d\bar r\|_{\bar r} = \|dr\|_r.
\end{equation}
We still need to demonstrate that the numerator is measured similarly. To achieve this, observe that the coordinate representation of the differential of the flow map transforms as (dependencies and indices dropped)
\begin{equation}
    \nabla\bar F = Q \nabla F Q^{-1}. 
\end{equation}
Consequently, the coordinate representation of the Cauchy--Green tensor transforms according to
\begin{equation}
    \bar C = (\nabla\bar F)^\top \bar G(\bar F)\nabla\bar F = Q^{-\top}(\nabla F)^\top Q^\top\bar G(\bar F)Q\nabla FQ^{-1} = Q^{-\top}CQ^{-1}.
\end{equation}
Then the numerator of \eqref{eq:L},
\begin{equation}
    \|d\bar F\|_{\bar F} = d\bar{\mathbf r}\cdot \bar C d\bar{\mathbf r} = d\mathbf r\cdot (Q^\top Q^{-\top}) C d\mathbf r =  d\mathbf r\cdot C  d\mathbf r =  d\mathbbl r\cdot \mathsf C  d\mathbbl r = \|dF\|_F,
\end{equation}
which is equally measured by the $x$- and $\bar{x}$-observers, thereby showing that \eqref{eq:L} is objective and all conclusions derived using it are independent of the observer's viewpoint.

\begin{remark}[On objectivity on $M = \mathbb R^2$]
    On $M = \mathbb R^2$ a general change of observer is defined by \textup{\cite[][Chapter 3]{Haller-23}} $\bar x = Q(t)x + b(t)$ for $Q(t) \in \operatorname{SO}(2)$ and $b(t) \in \mathbb R^2$ arbitrary.  The notion of objectivity follows as discussed above with $Q^t$ replaced by $Q(t)$, satisfying (with the $t$-dependence omitted) $Q^\top Q = \mathsf I = QQ^\top$ and $\det Q = 1$.
\end{remark}

\section{Birth and death framing in short-time 2-d flows}\label{sec:birthdeath}

An extended geodesic vortex detection iteration, as introduced in \cite{Andrade-etal-20}, used in \cite{Andrade-etal-22, Andrade-Beron-22}, and implemented in our study with a modification imposed by the nature of stratospheric flow, conducts a comprehensive exploration of the 2-d parameter space $(t_0,t_1)$, \emph{allowing one to frame birth and death of CLVs}. Specifically, the unmodified extension consists in iteratively applying geodesic vortex detection within the delineated flow domain as follows:
\begin{enumerate}[resume]
    \item Slide the initial time instance $t_0$ over a time window covering the time interval of during which a CLV is expected to exist.
    
    \item For each $t_0$, progress $T$ as long as a CLV boundary is successfully detected.  This way, for each $t_0$ a life expectancy $T_{\exp}(t_0)$ is obtained, which is the maximum $T$ for which a CLV over $[t_0,t_1]$ is successfully detected. The expected result is a linearly decaying, i.e., \emph{wedge-shaped}, $T_{\exp}(t_0)$, indicating that all Lagrangian coherence assessments predict the breakdown consistently, irrespective of any predetermined parameter settings.  The \emph{birth date}, $t_\mathrm{birth}$, of the vortex is given by $t_0$ for which $T_{\exp}(t_0)$ is maximized.  The \emph{death date}, is then $t_\mathrm{death} = t_\mathrm{birth} + T_{\exp}(t_\mathrm{birth})$.
\end{enumerate}
More precise evaluations of the birth and death dates of the CLV can be achieved by combining the outcomes derived from executing the algorithm in both forward- and backward-time directions \cite{Andrade-etal-20}.

The aforementioned modification of the birth-and-death CLV framing algorithm is dictated by the temporal scale $\tau$ within which the wind field is accurately depicted as a 2-d flow in the stratosphere. As noted above, after $\tau \approx 1$ month, air parcels initially positioned on a specific isentropic surface begin to undergo considerable dyapicnic mixing \cite{Haynes-05}. For rapidly transient phenomena, such as SSWs, where tracer distributions are predominantly affected by vertical motion, the 2-d model continues to serve as a reliable diagnostic instrument for evaluating horizontal mixing up to $\tau$ \cite{McIntyre-82}.

Consequently, the coherence horizon $T$ is restricted to ensure it does not exceed $\tau$. This constraint guarantees that $T_\text{exp}(t_0)$ remains within the limits of $\tau$. Within this constraint framework, Step 2, outlined earlier, is revised and divided in the following manner:

\begin{enumerate}
   \item[2.a)] It is anticipated that the theoretical $T_\text{exp}(t_0)$ will exhibit the form of a \emph{truncated} wedge at height $\tau$. The death date of the CLV can consequently be forecasted by determining the latest $t_0$, $\smash{t_0^\text{late}}$, for which $T_\text{exp}(t_0)$ maintains stability around $\tau$, augmented by $\tau$, viz., $t_\text{death} = \smash{t_0^\text{late}} + \tau$.

   \item[2.b)] To estimate the birth date of the CLV, it is necessary to apply the algorithm in backward time starting from $t_\text{death}$, as estimated above, moving in reverse chronological order. This involves using $0 \ge T \ge -\tau$ while regressing $t_0$ from $t_\text{death}$. Then, $t_\text{birth}$ will be determined as the earliest $t_0$, $\smash{t_0^\text{early}}$,  for which $|T_\text{exp}(t_0)|$ remains stable and close to $\tau$, minus $\tau$, i.e., $t_\text{death} = \smash{t_0^\text{early}} - \tau$.
\end{enumerate}

An important final observation is that we can no longer strictly refer to CLVs, but rather as quasi-CLVs. This is because the estimated $t_\text{death} - t_\text{birth}$ will exceed $\tau$, meaning that the CLV at $t = t_\text{death}$ will not be the advected image of the CLV at $t = t_\text{birth}$.

\section{The life cycle of the SVP in 2002}\label{sec:life}

\subsection{Velocity data and manifold parameterization}

In this study, we have elected to examine isentropic winds derived from reanalysis data. 
A reanalysis system synthesizes historical short-term forecasts with observational data through the process of data assimilation.The reanalysis framework under consideration is the \href{https://www.ecmwf.int/en/forecasts/dataset/ecmwf-reanalysis-v5}{European Centre for Medium-Range Weather Forecasts (ECMWF) Reanalysis v5 (ERA5)}, which is the latest ECMWF reanalysis available \cite{Hersbach-etal-20}. We have selected the 600\,K isentropic surface, which is positioned well within the stratosphere. The wind fields examined are restricted to the Southern Hemisphere for the year 2002 and are presented with a horizontal resolution of 31 km, adhering to an eighth-day temporal resolution. Daily averaged velocities have been employed in this analysis to speed up computations and mitigate the numerical noise that is intrinsic to assimilation systems. The primarily spectrally nonlocal character of mixing within the stratosphere \cite{Shepherd-etal-00} ensures that the implemented filtering does not significantly alter Lagrangian transport evaluations.

Given that our objective is to conduct geodesic vortex detection near the southern pole, we have opted to avoid using geographic longitude--latitude coordinates  $(\lambda,\vartheta) \in [0,2\pi]\times [-\frac{\pi}{2},\frac{\pi}{2}]$ due to their singularity at this location. Instead, we have chosen to parameterize the southern hemispherical cap $S^2_-$, the target manifold $M$, following \cite{Lekien-Ross-10}. This entails positioning the ambient Euclidean space $\mathbb R^3$ with coordinates $(x,y,z)$, such that $(x,y)$ aligns with the equatorial plane, and $z$ is oriented in the northward direction. Then one writes
\begin{equation}
    \varphi(x) =
    \begin{pmatrix}
        x^1\\ x^2\\ -\sqrt{a^2 - (x^1)^2 - (x^2)^2} 
    \end{pmatrix}
    \label{eq:xy}
\end{equation}
for $(x^1)^2 + (x^2)^2 > a^2$, where $a \approx 6387$\,km is the arithmetic mean of the Earth's radius combined with the altitude of the stratosphere's base. The coordinate representation of the $\varphi$-induced pullback metric on $M$ is given by
\begin{equation}
    G(x) = 
    \begin{pmatrix}
        \frac{a^2 - (x^2)^2}{a^2 - (x^1)^2 - (x^2)^2} & \frac{x^1x^2}{a^2 - (x^1)^2 - (x^2)^2}\\
        \frac{x^1x^2}{a^2 - (x^1)^2 - (x^2)^2} &  \frac{a^2 - (x^1)^2}{a^2 - (x^1)^2 - (x^2)^2}
    \end{pmatrix}.
    \label{eq:metric}
\end{equation}
Given longitudinal, $u_\lambda(\lambda,\vartheta,t)$, and meridional, $u_\vartheta(\lambda,\vartheta,t)$, wind components as retrieved from the ERA5 reanalysis data base on the 600-K isentropic surface in the Southern Hemisphere, we transform $(\lambda,\vartheta) \mapsto (x^1,x^2)$ according to
\begin{equation}
    x^1 = a\cos\vartheta\cos\lambda,\quad 
    x^2 = a\cos\vartheta\sin\lambda,
\end{equation}
and apply geodesic vortex detection on trajectories numerically generated by
\begin{equation}
    \left\{
    \begin{aligned}
        \dot x^1 &= - u_\lambda \sin\lambda - u_\vartheta\sin\vartheta\cos\lambda = u^1(x^1,x^2,t),\\
        \dot x^2 &= + u_\lambda \cos\lambda - u_\vartheta\sin\vartheta\sin\lambda = u^2(x^1,x^2,t).
   \end{aligned}
   \right.
   \label{eq:dxdt1}
\end{equation}
Although the chosen coordinate system for $S^2_-$ is free from singularity at the southern pole ($x = 0$) unlike spherical coordinates (compare \mbox{\eqref{eq:dxdt1}} with $\dot\lambda = u_\lambda/a\cos\vartheta$, $\dot\vartheta = u_\vartheta/a$), it is not orthogonal, as the nondiagonal nature of $G(x)$ in \mbox{\eqref{eq:metric}} indicates. However, this does not present any mathematical or computational challenges.

\subsection{Numerical implementation of geodesic vortex detection}

Our computational framework of choice is the \href{https://julialang.org/}{Julia} package \href{https://github.com/CoherentStructures/CoherentStructures.jl}{CoherentStructures.jl}, which has been modified to operate on flow defined on a 2-d manifold in arbitrary coordinates, as detailed in the previous section. A minimal working script has been made accessible to the public as detailed in the Data Availability section below. The algorithm is reviewed in Appendix \ref{app:num}. For our applications, we employed a numerical grid comprising $256 \times 256$ points. The chosen integration scheme was the Tsitouras \cite{Tsitouras-11} method of the 5(4)th-order Runge–Kutta family.

\begin{figure}[t!]
    \centering  
    \includegraphics[width=.2\textwidth]{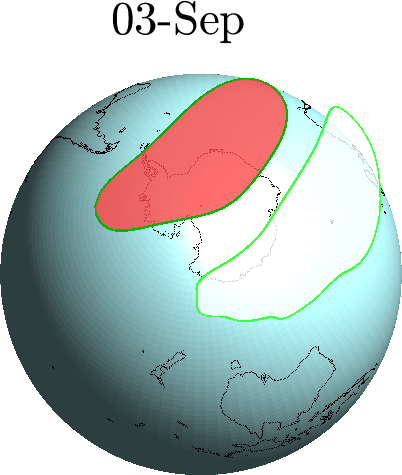}\,
    \includegraphics[width=.2\textwidth]{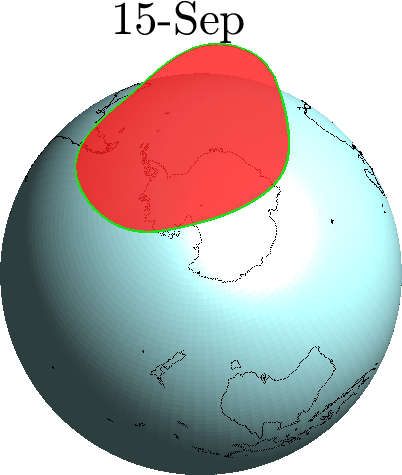}\,
    \includegraphics[width=.2\textwidth]{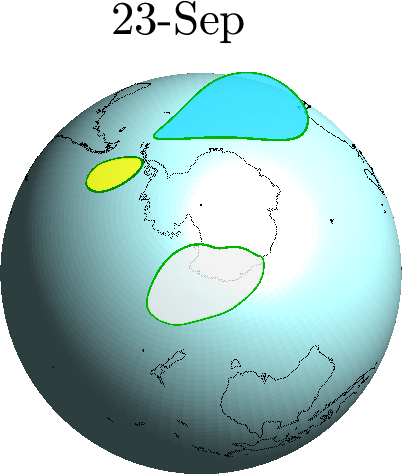}\,
    \includegraphics[width=.2\textwidth]{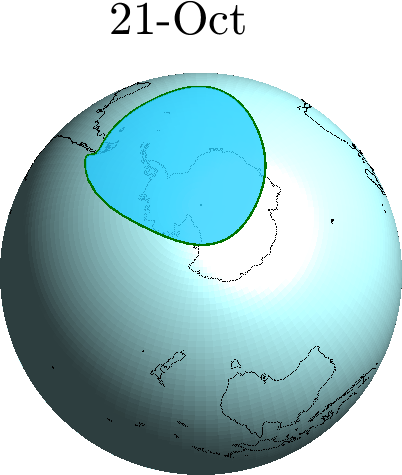}\\\bigskip
    \includegraphics[width=.2\textwidth]{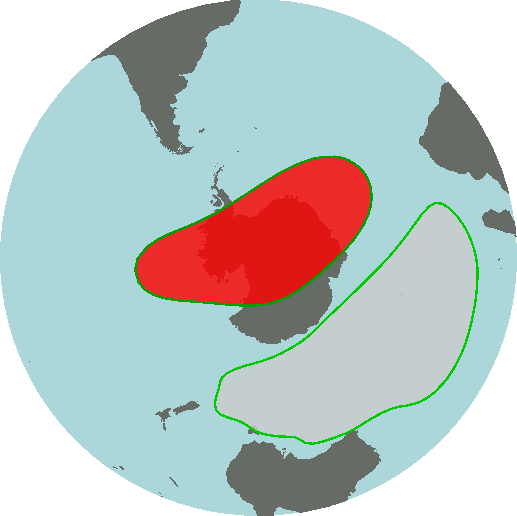}\,
    \includegraphics[width=.2\textwidth]{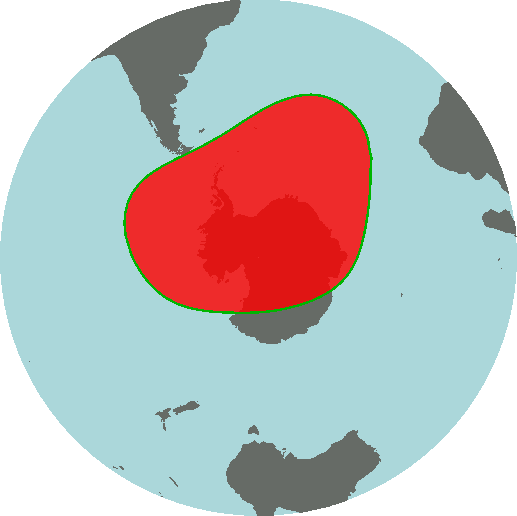}\,
    \includegraphics[width=.2\textwidth]{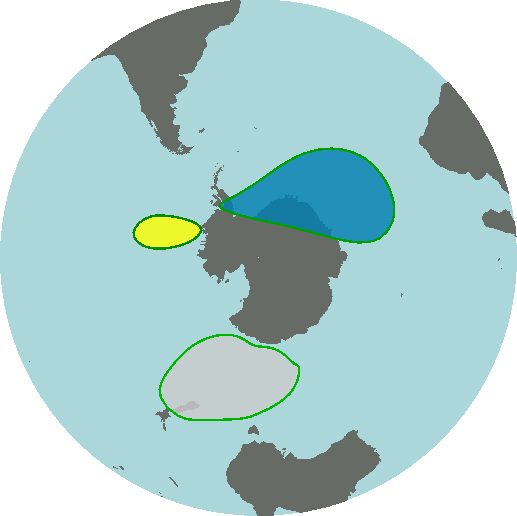}\,
    \includegraphics[width=.2\textwidth]{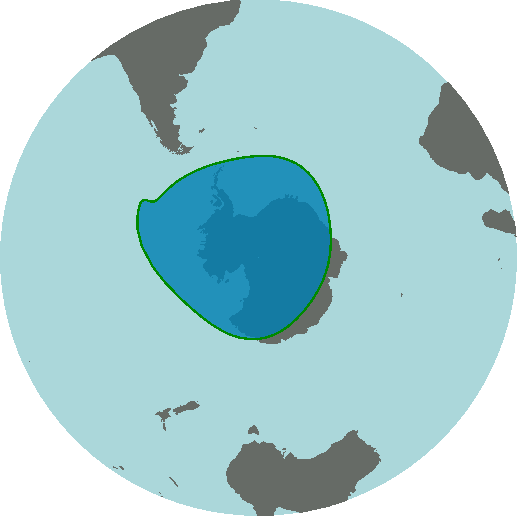}\\\bigskip
    \includegraphics[width=.65\textwidth]{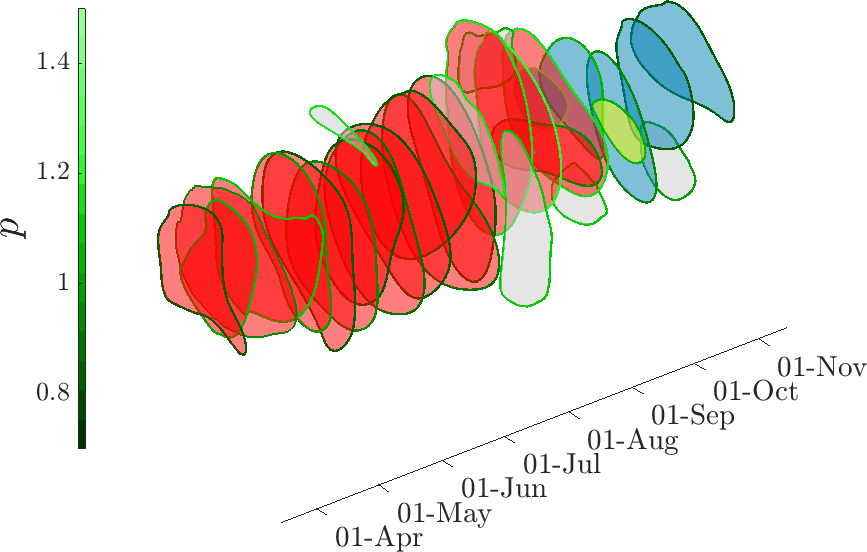}
    \caption{The results show the application of geodesic vortex detection to reanalyzed winds on the 600-K isentropic surface over time windows $[t_0,t_0+T]$, with $t_0$ rolling throughout 2002 and $T = 10$ days. Each detected CLV boundary is colored based on its stretching value ($p$). The interiors of the CLVs, which can be identified with different stages of the SPV due to their persistence, are shown in red prior to the SPV's splitting caused by the SSW in 2002, and in blue and yellow after the splitting. Transient CLV interiors are displayed in light gray. Acronyms used are defined in the main text and conveniently listed in Table \ref{tab:acro}.}
	 \label{fig:roll}
\end{figure}

\subsection{Quick, rolling time window analysis}

We begin by applying geodesic vortex detection over time windows $[t_0, t_0+T]$, with a Lagrangian coherence horizon $T = 10$ days, while rolling over the Lagrangian coherence assessment time $t_0$ throughout 2002 with a forward-time step $\Delta t_0 = 10$ days. This methodology facilitated a rapid and computationally efficient, albeit approximate, characterization of the birth and death of the austral SPV classified as an CLV, as well as the splitting event induced by the extraordinary SSW experienced by the SVP in 2002, alongside any other CLVs that happened to be geodesically detected.  The selection of $T = 10$ days is based on an approximate estimate of the persistence of the smaller vortex resulting from the splitting of the SPV. Although this choice facilitates the delineation of the vortex in question, it exemplifies the kind of parameter preset we aim to circumvent, as will be demonstrated in the subsequent section.

The results of the proposed quick, rolling time window analysis are shown in Figure \ref{fig:roll}.  The top and middle panels illustrate multiple boundary extractions of the CLV. In particular, the upper panels display these boundaries embedded within the ambient Euclidean space according to the parameterization $\mathbb R^3 \hookleftarrow M\ni (x^1,x^2) \mapsto \varphi(x^1,x^2) \in \mathbb R^3$ defined by \eqref{eq:xy}, with $a$ exaggerated.  In contrast, the lower panels represent these boundaries as mapped on a plane tangent to the South Pole via stereographic projection.  These highlight the splitting phenomenon of the CLV, which can be identified with the SPV. We stress that this identification is only approximate, as the air regions contained within the extracted CLV boundaries depicted are not advected images of one another under the isentropic flow. Yet, the repeated emergence of the CLVs over time suggests that they are related. With this in mind, three different colors were chosen to distinguish the stages of the SPV before and after the splitting. Specifically, red is used to depict the vortex interior before the splitting. Blue and yellow are employed to show the most and less persistent of the two vortices into which the SPV splits, respectively. These colors more precisely depict the area of air surrounded by the SPV's edge, colored at each $t_0$ according to the stretching parameter ($p$) over $[t_0, t_0+T]$. The bottom panels depict additional CLV extractions across various distinct $t_0$ within the space defined by coordinates $(x^1,x^2)$, extended over the coherence assessment time. Note that, consistent with the aforementioned comments on the persistence of the identified SPV, the parameter $p$ for the edge(s) of the SPV(s) prior to the splitting remains close to 1. Similarly, the SPV offspring depicted in blue remains nearly constant at a parameter $p$ close to 1, but the smaller SPV offspring shown in yellow cannot be traced beyond approximately 7 days, indicating a much shorter lifespan.  That $p$ is close to 1 is consistent with the material coherence assessment being restricted to only 10 days, in principle.

The geodesic vortex detection algorithm discerns multiple CLVs, in addition to those that can be associated with varying phases of the SPV. The interiors of these vortices are illustrated in Fig.\@~\ref{fig:roll} in light gray. These CLVs are characterized by a relative stretching of up to $p \approx 1.3$, are significantly shorter in duration, emerging and intermittently vanishing, which precludes their temporal correlation. Certain transient vortices, particularly that begin in spring, represent manifestations of the equatoward breaking of upward-propagating Rossby waves \cite{Ngan-Shepherd-99a}, which theoretically organize into a cat-eye configuration within critical layers formed along lines where the zonal wind vanishes \cite{Stewartson-78, Warn-Warn-78}.

It should be noted that the calculation of elliptic objective Eulerian coherent structures \cite{Serra-Haller-16}, adapted for curved surfaces, is anticipated to offer an assessment similar to the one given above. These structures differ from those discussed above because they relate to the limit $T \to 0$, which may result in even further computational savings. Furthermore, they include a factor similar to $p$, but interpreted as an instantaneous material stretching rate.

\subsection{Detailed birth-and-death framing}

This section discusses the birth and death of the SPV in 2002, implementing the algorithm described in Section \ref{sec:birthdeath}. Ideally, the algorithm would be applied from flexible early and late start and end points of the Lagrangian coherence assessment time $t_0$. However, the SPV's division limits this flexibility. Therefore, our analysis of birth and death focuses on the SPV before the split and afterward, involving the birth and death of the two vortices resulting from the SPV's division.  

In a broader context, the occurrence of multiple vortices emerging and vanishing at various times within a specified interval makes it impractical to conduct a birth-and-death analysis autonomously. The rolling time window analysis from the preceding section is crucial in providing guidance to effectively execute the analysis.

\begin{remark}\label{rem:split}
    The concept of splitting, within deterministic dynamical systems, is related to the transport described by the fluid particle equation \eqref{eq:dxdt} and assumes a smooth velocity field and a unique flow map \eqref{eq:F}. It does not imply an actual splitting, but refers to the process in which a fluid region is stretched and folded, eventually appearing as two regions connected by thin filaments without true separation.
\end{remark}

\subsubsection{Analysis prior the splitting event}

\begin{figure}[t!]
    \centering
    \includegraphics[width=.9\linewidth]{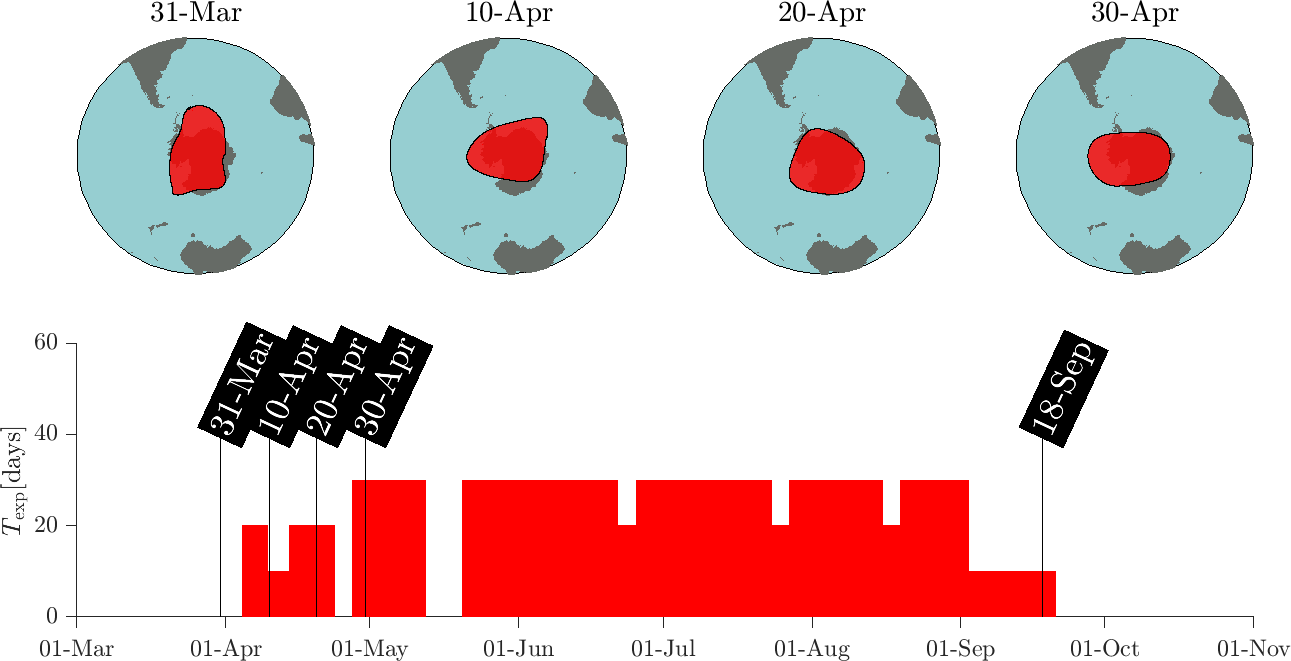}
    \caption{The birth of the austral SVP in 2002 depicted through geodesic vortex detection. The top panels demonstrate various stages of the formation process, while the bottom panel depicts the life expectancy against Lagrangian coherence time, as estimated numerically in a backward direction. The stretching parameter varies from $p \approx 0.94$ to 1.12.}  
    \label{fig:birth}
\end{figure}

We begin by characterizing the formation of the SPV before splitting. From Figure \ref{fig:roll}, we see that the appropriate endpoint for analysis is $t_0 = 25$ September 2002. We then conduct geodesic vortex detection backward in time with $T = - \tau$, with $\tau = 30$ days. As discussed in Section \ref{sec:birthdeath}, the Lagrangian coherence assessment date $t_0$ is moved backward from the selected endpoint in $\Delta t_0 = -5$ day steps while maintaining the Lagrangian coherence horizon at $T \le \tau$. The results are in the lower panel of Figure \ref{fig:birth}. The algorithm remains irresponsive to CLV detection until $t_0 = 18$ September 2002. Tracing back to $t_0^\text{early} = 30$ April 2002, the algorithm consistently detects a CLV with $T = -\tau$. The observed variability is anticipated as a consequence of the intrinsic noise present in reanalyzed velocity data. On $t_0^\text{early} = 30$ April 2002, the life expectancy $T_\text{exp}(t_0)$ starts a near-linear decline, and there is no CLV detection prior to $t_0 = 31$ March 2002. This evidences, within the limits of numerical noise, a truncated wedge for $T_\text{exp}(t_0)$ consistent with theoretical expectation, validating the predicted birth date $t_\text{birth} = t_0^\text{early} - \tau = 31$ March 2002. Several stages of the 2002 SPV birth are shown in Figure \ref{fig:birth} (top panels). These panels depict, from right to left, backward-advected CLV images detected on $t_0^\text{early}$; the leftmost stage marks the SPV birth as verified by geodesic vortex detection.

\begin{figure}[t!]
    \centering
    \includegraphics[width=.9\linewidth]{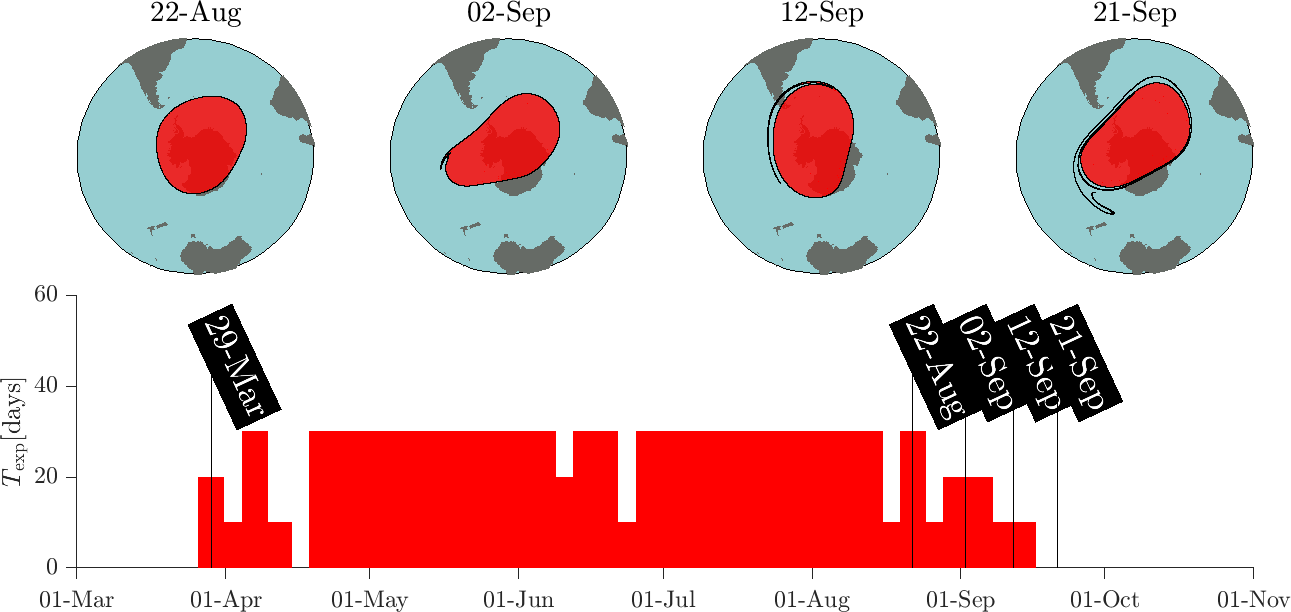}
    \caption{As in Figure \ref{fig:birth}, but computed in forward time to frame the death of the SPV prior the splitting.  The stretching parameter of the extracted CLV boundaries varies from $p \approx 0.93$ to 1.2.}
    \label{fig:death}
\end{figure}

The bottom panel of Figure \ref{fig:death} shows life expectancy as a function of the Lagrangian coherence assessment time, $T_\text{exp}(t_0)$, calculated in the forward direction. This is achieved using $T = \tau$ and employing the sliding $t_0$ approach starting from $t_0 = 23$ March 2002, with $\Delta t_0 = 5$ day. Geodesic vortex detection identifies a CLV for the first time on $t_0 = 29$ March 2002. Changing the starting $t_0$ to an earlier date in 2002 does not alter this inference. After $t_0 = 26$ March 2002, $T_\text{exp}(t_0)$ remains relatively stable with some noise around $\tau$ until $t_0^\text{late} = 22$ August 2002. From that date onward, $T_\text{exp}(t_0)$ decays fairly linearly until $t_0 = 12$ September 2002. The hypothesized truncated wedge form for $T_\text{exp}(t_0)$ becomes apparent when the birth-and-death algorithm is applied in a forward manner, akin to its application in a reversed time order, subject to the influence of numerical noise. The death date is calculated to be $t_\text{death} = t_0^\text{late} + \tau = 21$ September 2002. The forward-advected images of the SVP as geodesically detected on $t_0^\text{late}$ in the top panels effectively corroborate the assessment of the death date of the vortex before splitting.

It is important to note that discrepancies in birth-and-death date assessments arise between forward- and backward-time analyses. Specifically, a death date identified by forward-time analysis can be perceived as more precise compared to a backward-time evaluation, while the converse holds true for birth date evaluations \cite{Andrade-etal-20}. Such variations are reflected in the calculations reported above. For example, the backward-time analysis places the date of death at 18 September 2002, which is 3 days prior to the date identified by the forward-time analysis, noted for its greater accuracy. However, the backward-time analysis estimates the birth date as 31 March 2002, which is only 2 days later than the date determined by the less precise forward-time evaluation. These observed differences, particularly regarding the evaluation of the death date, are not deemed significant given the extended lifespan of the SPV prior to splitting, which is close to 180 days.

\subsubsection{Analysis past the splitting event}

\begin{figure}[t!]
    \centering
    \includegraphics[width=.75\linewidth]{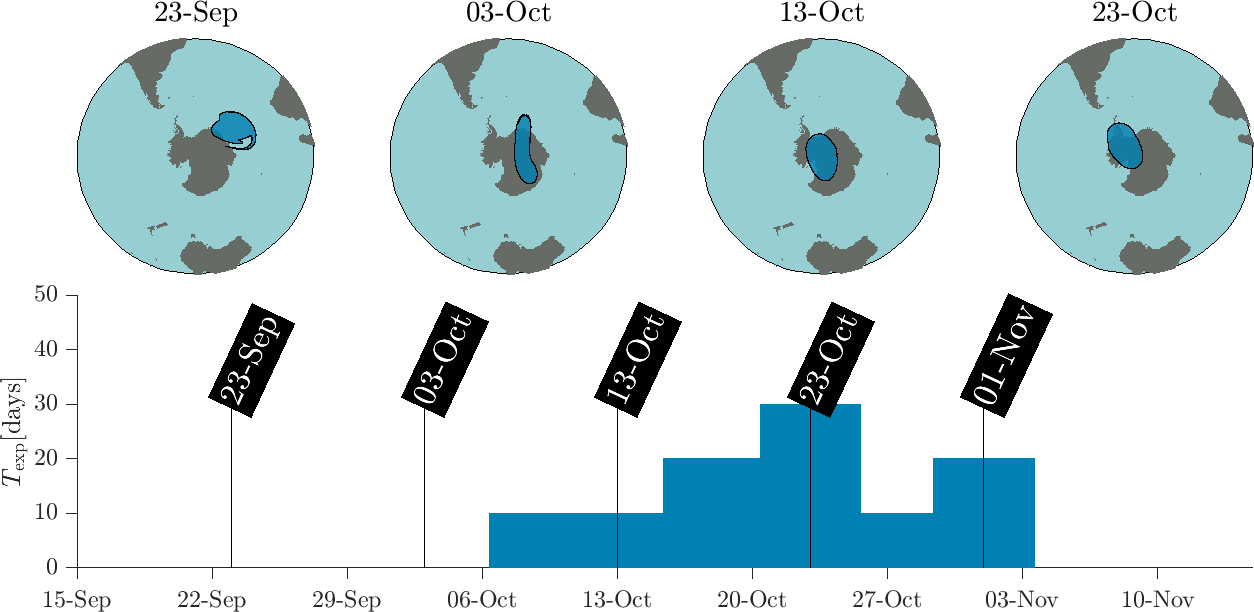}\\\bigskip
    \includegraphics[width=.75\linewidth]{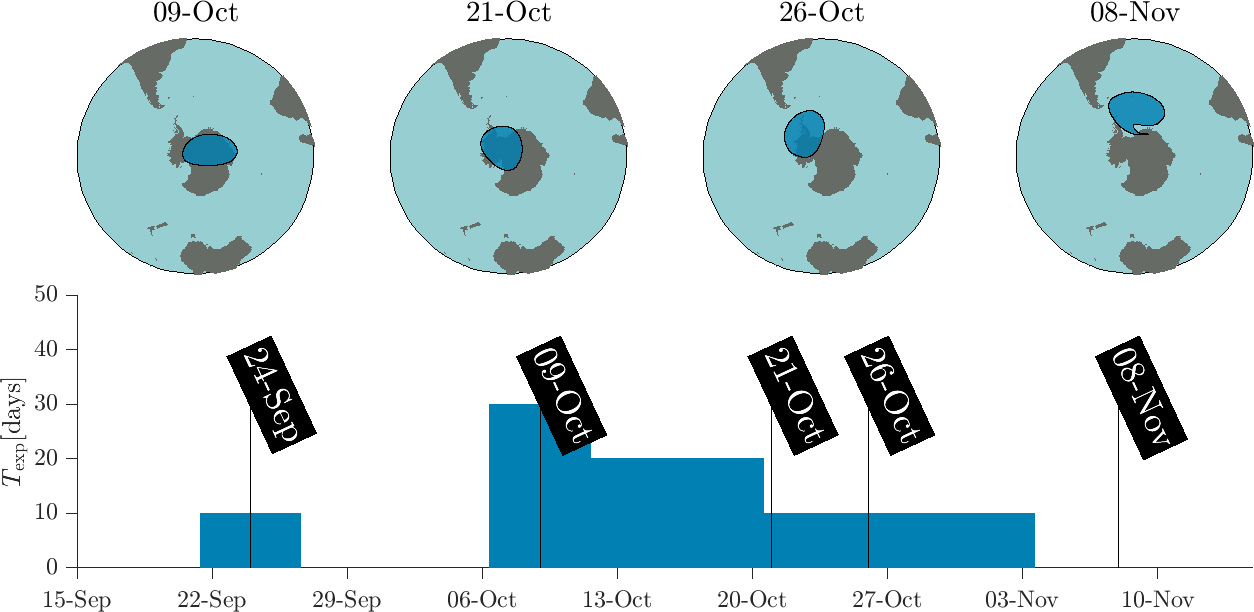}
    \caption{The top and bottom panels show the same information as Figures \ref{fig:birth} and \ref{fig:death}, respectively, but for the larger of the two vortices formed when the austral SPV split in 2002 as a result of the SSW. The stretching parameter varies from $p\approx 0.92$ to 1.18 in the top, and from $p\approx 0.86$ to 1.18 in the bottom.}
    \label{fig:blue}
\end{figure}

Following the splitting of the austral SVP event produced by the SSW in 2002, the birth-and-death CLV framing algorithm identifies two vortices with varying durability. The more enduring and larger vortex is determined to be born on $t_0^\text{birth} = 23$ September 2002 and die on $t_0^\text{death} = 8$ November 2002, through the application of the algorithm in backward and forward temporal directions, respectively. The corresponding estimates $T_\text{exp}(t_0)$ are shown in Figure \ref{fig:blue} together with backward (top) and forward (bottom) advected images of the CLV as identified on $t_0^\text{early} = 23$ October 2002 and $t_0^\text{late} = 09$ October 2002. It should be noted that, given that the lifespan of this fragmented vortex is approximately 30 days, $T_\text{exp}(t_0)$ does not present a truncated wedge shape but rather exhibits a more complete wedge shape, accounting for numerical inaccuracies corresponding to the instability of this vortex compared to that preceding the splitting event.  Consistent with the analysis conducted prior to the splitting event, anticipated discrepancies emerge in the assessments of the birth-and-death dates when employing the birth-and-death framing in forward and backward temporal analyses. The forward-time analysis identifies the birth of the relevant CLV as occurring on 9 October 2002, which is just 1 day subsequent to the date identified by the backward-time analysis. On the other hand, the death date is determined by the backward-time analysis as 1 November 2002, approximately 7 days prior to the date determined via forward-time evaluation.

\begin{figure}[t!]
    \centering
    \includegraphics[width=.75\linewidth]{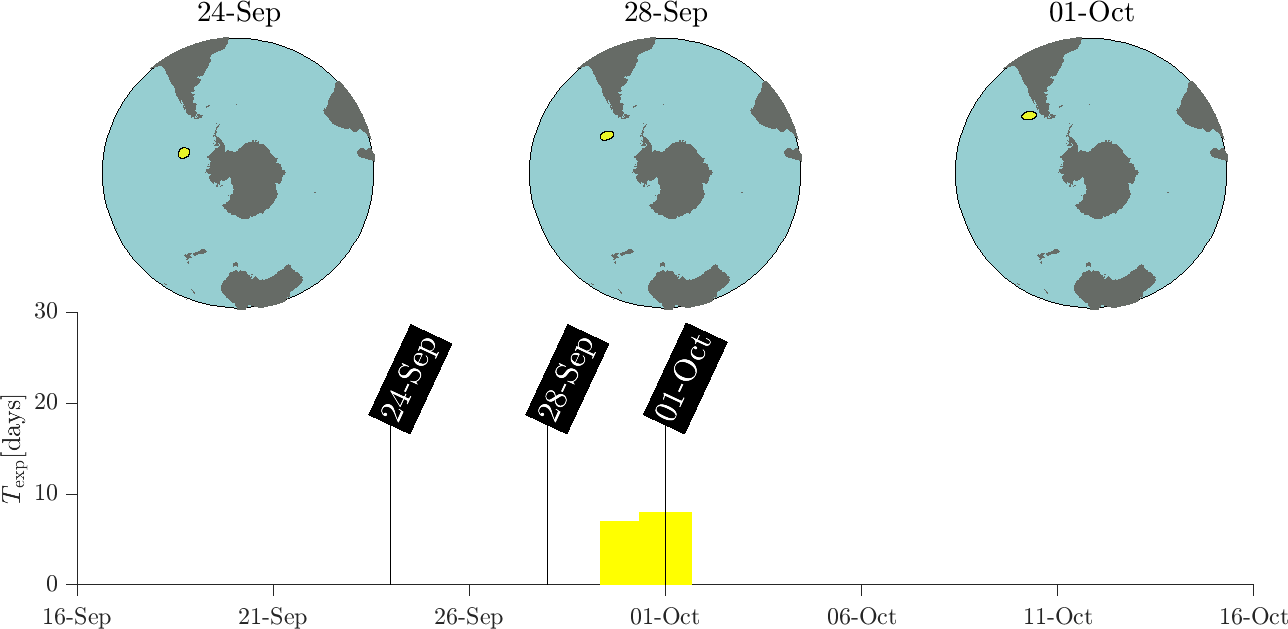}\\\bigskip
    \includegraphics[width=.75\linewidth]{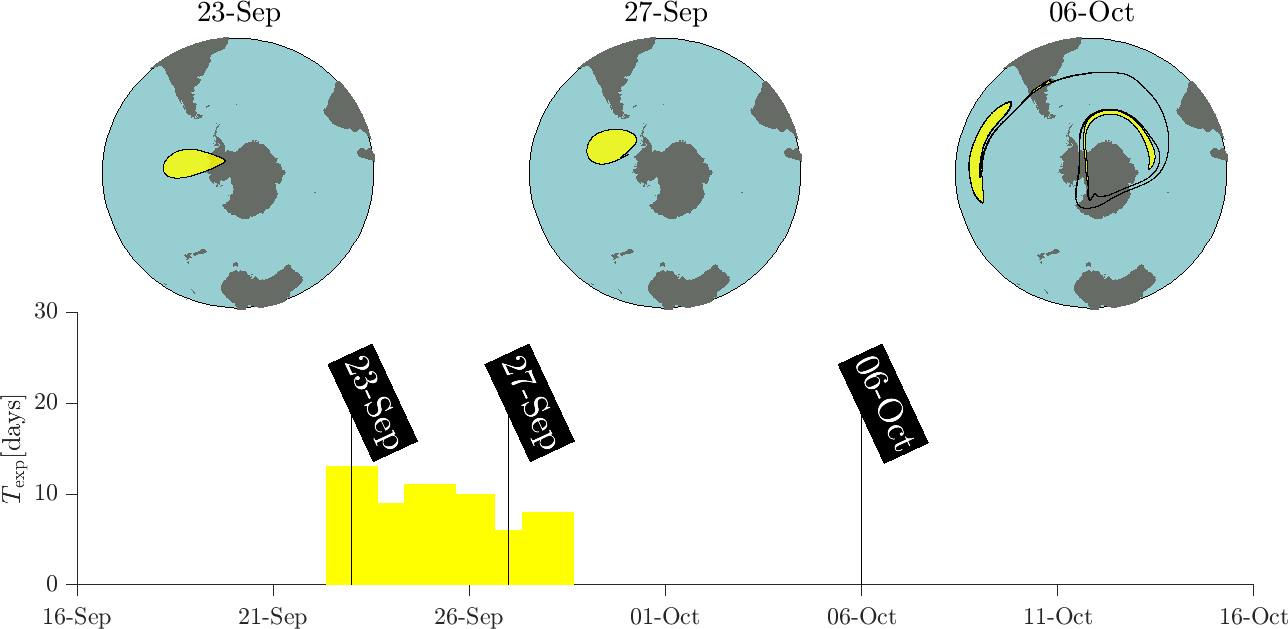}
    \caption{Similar to Figure \ref{fig:blue}, this illustration pertains to the smaller of the two vortices that emerge from the splitting process. The stretching parameter varies from $p\approx 0.93$ to 1.17 in the top, and from $p\approx 0.86$ to 1.1 in the bottom.}
    \label{fig:yellow}
\end{figure}

The smaller of the two vortices resulting from the splitting exhibits a shorter lifespan compared to the larger vortex. Both forward and backward-time birth-and-death CLV framings estimate a lifespan not exceeding 20 days. Consequently, the coherence assessment period $t_0$ was adjusted with a time step of $\Delta t_0 = 1$ days to facilitate a more precise determination of the birth and death dates. The birth date, based on analyses conducted in both temporal directions, is estimated to occur between 23 and 24 September 2002. In a similar vein, the death date is more consistently assessed to fall between 1 and 6 October 2002. The curve patterns expected for $T_\text{exp}(t_0)$ appear with significant computational errors, possibly due to the short duration of the CLV and the inherent noise in the reanalyzed isentropic winds, as shown in Figure \ref{fig:yellow}.

\subsubsection{Assessing the splitting instant}

Using the results of the birth-and-death CLV framing analysis conducted before and after the splitting of the SPV, it becomes feasible to determine the moment at which this phenomenon occurs. Consistent with the CLV notion, and the methodology used to detect CLVs, it is appropriate to designate this moment as the earliest time $t_0$ when the resulting products of the splitting, specifically the two CLVs into which the SPV splits into, are simultaneously evaluated as being alive, according to geodesic vortex detection. This point $t_0$ can be established, by comparing the top panels of Figure \ref{fig:blue} with those of Figure \ref{fig:yellow}, on 23 September 2002.

\begin{figure}[t!]
    \centering
    \includegraphics[width=\linewidth]{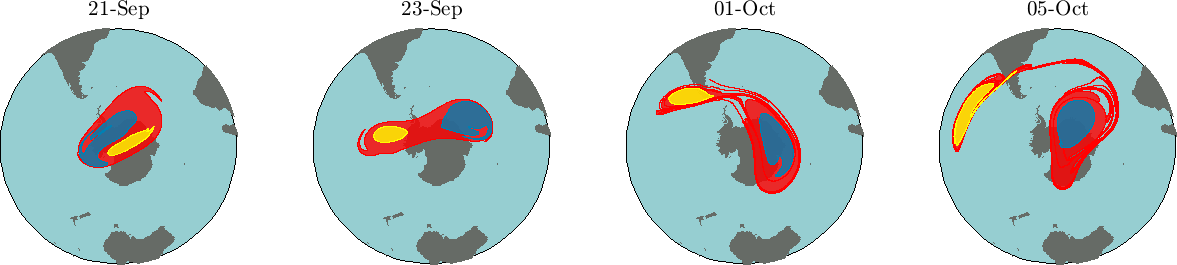}
    \caption{Illustration of the splitting of the SPV resulting from the SSW in 2002 as framed by geodesic vortex detection.}
    \label{fig:split}
\end{figure}

While the proposed criterion suggests 23 September 2002 as the splitting date, the birth-and-death CLV framing algorithm indicates that the SPV died on 21 September 2002, prior to the split; cf.\ Fig.\@~\ref{fig:death}. This minor discrepancy adds credibility to the proposed criterion. However, the split of the SPV due to the SSW in 2002 is a gradual process. This phenomenon is clearly visible in Fig.\@~\ref{fig:split}, showing advected images of the geodesically detected CLVs on the inferred splitting date 23 September 2002. The images on the left are from backward advection, while those on the right are from forward advection. Initially, notice the genesis of the split SPV species identified geodesically within the SPV, impacting the trapping of ozone-depleted air as discussed in Section~\ref{sec:o3}. However, it is not until 1 October 2002 that the complete split becomes visually clear. The SPV maintains some material coherence beyond the geodesically inferred splitting date, on 23 September 2002. During this period, and at least until 5 October 2002, the geodesically identified split pieces of the SPV remain within the air mass outlined by its material boundary, which eventually undergoes significant stretching. The final phase of the SPV life cycle is primarily influenced by the largest of the two CLVs, into which the main CLV splits.

\section{Ozone hole kinematics}\label{sec:o3}

\begin{figure}[t!]
    \centering
    \includegraphics[width=\linewidth]{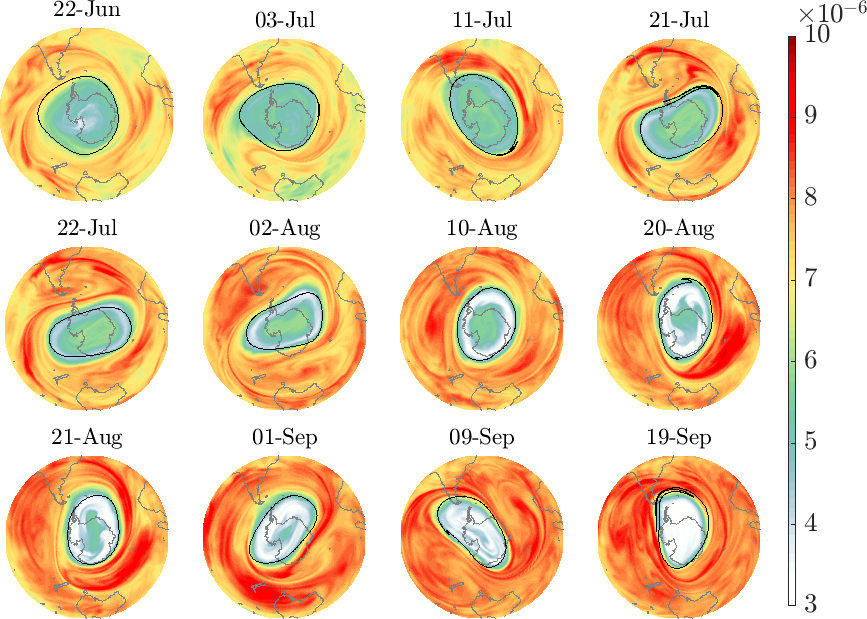}
    \caption{Overlaid on the ozone mass fraction per kilogram of air is the edge of the SPV (thick black loop) as extracted using geodesic vortex detection from reanalyzed isentropic winds. The geodesic vortex detection is applied $T = 30$ days forward from the Lagrangian coherence assessment time $t_0$ as shown in the leftmost panel of each row. The other panels in each row depict advected images of the computed CLV boundary.}
    \label{fig:o3}
\end{figure}

An independent examination of the transport barrier characteristics of the SPV edge, identified using geodesic vortex detection on the wind field, is shown in Figure \ref{fig:o3}. This analysis uses ozone concentration data from the ERA5 reanalysis system. Although the wind field used for geodesic vortex detection was generated by the same system, it cannot be assumed in advance that the SPV Lagrangian boundary derived from geodesic methods will effectively confine ozone-depleted air. It is expected to do so with high accuracy.  This has been previously observed \cite{Serra-etal-17}, predicated on the assumption that isentropic surfaces are planar. Here, we perform this analysis not only for completeness, but also to shift our focus toward the kinematics of ozone depletion.

We thus performed geodesic vortex extractions on $t_0 = $ 22 June,  22 july  and 21 August 2002 with $T = 30$ days, nearing the maximum time air parcels can be expected to remain on the selected isentropic surface of 600~K. The extracted Lagrangian edge of the SPV is shown in the first panel of each row of Figure \ref{fig:o3}; subsequent panels correspond to advected images of this material loop on selected dates. Observe the coherence of the material loop, which resists unrestricted stretching as anticipated. While this inference holds true subject to numerical error, it becomes less evident in September, prior to the splitting event. Indeed, in each instance, the material loop that constitutes the SPV edge at time $t_0$ is a limit cycle of a $p$-line field \eqref{eq:p-line} with $p \approx 0.9$. 

As noted, the stretched state of the advected image of the above $p$-loop on $t_0 + T = 19$ September 2002 is challenging to reconcile with the CLV assessment on $t_0 = 21$ August 2002. An explicit calculation of the relative stretching experienced by this material loop over $[t_0,t_0+T]$ reveals a value of about 2.15. Although computationally rigorous, this assessment is far less consistent with the absence of deformation within the CLV enclosed by the material loop. In fact, the stretching of the loop is predominantly tangential. The explanation for this phenomenon is that the material line in question does not represent a perfect material loop, allowing for flow-shear-induced stretching. Increasing the resolution of the computational grid used for geodesic detection did not mitigate this phenomenon, which we attribute to noise in the realized wind field, likely amplified during the SSW event.

The extracted material loops are superimposed on the ozone mass mixing ratio, specifically the mass fraction of ozone per kilogram of air. It is important to note that ozone destruction begins within an annular region defined by the geodesically detected boundary of the SPV and subsequently moves inward toward the SPV's center. This serves as a robust test of the transport barrier property of the edge for passive tracers, as chemical species like ozone are expected to behave approximately.

The evolution of ozone depletion is as anticipated; however, it is frequently assessed relative to the edge of the SPV, as defined by potential vorticity arguments \cite{Nash-etal-96}. It should be noted that potential vorticity depends on the observer's perspective, as shown in the Online Supporting Information of \cite{Andrade-Beron-22}.  This affects any deductions regarding the edge of the SPV.  Our study provides objective (i.e., observer-independent) support to the previously established notion \cite{Solomon-99} that the poleward side vicinity of the edge of the SPV serves as the optimal zone for the formation and accumulation of polar stratospheric clouds (PSCs), due to the typically colder and more humid atmospheric conditions in this region. These PSCs facilitate photochemical reactions that result in ozone depletion.  Such phenomena typically occur in early spring, when atmospheric temperatures remain adequately low, yet sufficient solar irradiance is present. The SSW event in 2002 precipitated an earlier onset of ozone depletion. 

\begin{figure}
    \centering
    \includegraphics[width=\linewidth]{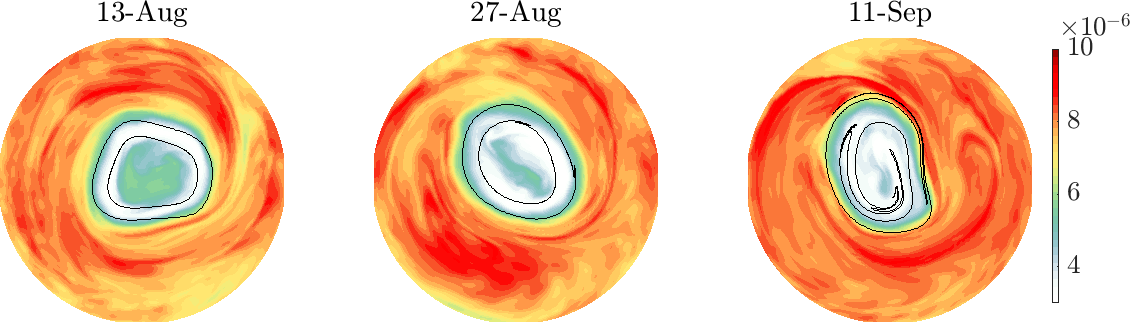}
    \caption{Similar to Figure \ref{fig:o3}, this figure includes the $p$-loop located nearest to the poleward border of the low-ozone concentration ring on August 12, 2002, and excludes continental borders.}
    \label{fig:o3-ring}
\end{figure}

A specialized geodesic vortex detection implementation provides deeper kinematic insights into ozone depletion within the SPV. The edge of the SPV prevents ozone-depleted air from spreading equatorward to mix with ozone-rich air. \emph{Another transport barrier, specifically an elliptic LCS located poleward of the SPV's edge, must be providing the necessary isolation for PSC formation and subsequent ozone depletion, temporarily, at least.} Examination of Figure \ref{fig:o3}, especially the middle and bottom rows, indicates that ozone-poor air in that annular region eventually spreads into the SPV's interior. In seeking a shorter-lived elliptic LCS, we applied geodesic vortex detection on $t_0 = 13$ August 2002, with $t = 30$ days, when the low-ozone ring was well-developed. Figure \ref{fig:o3-ring} shows the results, depicting the boundary of the SPV as the outermost $p$-loop, determined through an analysis of $p$-values across progressively expanding neighborhoods of 1.  This $p$-loop has $p \approx 1.05$. Its advected images over $[t_0, t_0+T]$ show that the SPV's edge, classified as a CLV, stretches more than theoretically predicted.  But this happens mainly in a tangential manner as it has been already argued. In addition to the SPV's edge, on $t_0 = 13$ August 2002, we extracted the $p$-loop closest to the poleward boundary of the low-ozone concentration ring on that day. This $p$-loop has $p \approx 1.6$, and by $t_0+T = 11$ September 2002, it shows visible signs inward filamentation. Consistent with the presence of the two $p$-loops extracted, the ozone-poor air is not mixed equatorward, but rather poleward. The poleward boundary of the low-ozone ring does not precisely follow the movement of the (synthetic) air parcels forming the 1.6-loop. Diffusion and reaction play significant roles, facilitated by the lack of a robust transport barrier like the SPV's edge. It is noted that the enhanced resilience of the material edge of the SPV, as identified geodesically, compared to the inner $p$-loop, can be attributed to its closer proximity to the core of the polar night jet, which is robust under perturbations, as discussed in Section \ref{sec:spvs}.

An important consideration is the reliability of the results described previously. The main concern is the quality of the stratospheric flow representations, particularly within the SPV, in the ECMWF ERA5 reanalysis system, which is affected by limited observations in the stratosphere compared to the troposphere and the scarcity of reliable anchoring data sets available \cite{Shepherd-etal-18}. Meanwhile, the ECMWF Integrated Forecasting System (IFS) includes a simplified representation of ozone chemistry, which incorporates the chemical processes responsible for the ozone hole \cite{Polichtchouk-etal-21}. However, the ERA5 reanalysis lacks support from satellite-derived ozone concentration data for the region around the South Pole and much of Antarctica until late September 2002.

\section{Summary}\label{sec:conclusion}

In this study, we have advanced a technique in nonlinear dynamical systems to identify vortices that exhibit flow-invariant or material boundaries, temporarily resisting the conventional exponential stretching encountered by arbitrary material loops in 2-d turbulent flow. This advancement is inspired by the intrinsic properties of natural flows and limitations regarding the temporal validity of the 2-d assumption.

The method in question is designated as ``geodesic vortex detection,'' which delineates vortices with material boundaries that extremize integrated relative stretching, evaluated objectively or in an observer-independent fashion. The solutions are (null) closed geodesics of a generalized Green--Lagrangian tensor, thus motivating the terminology, characterized by the property of uniform stretching, by an amount $p$. These geodesics are determined by limit cycles of a line field constituted by a linear combination of the eigenvectors of the coordinate representation of the (right) Cauchy–Green strain tensor for flows on the Euclidean plane in Cartesian coordinates, with coefficients dependent on the corresponding eigenvalues and $p$. Given that such a $p$-line field locally manifests as a rotated line field where it is orientable, its limit cycles are nonintersecting, and the outermost of a family of ``$p$-loops'' serves as the boundary of a coherent Lagrangian vortex.  

Our methodological advancement involves the extension of geodesic vortex detection to flows on 2-d Riemannian manifolds in arbitrary coordinates. This required the initial construction of an appropriate $p$-line field, wherein the eigenvectors and eigenvalues are derived not from the coordinate representation matrix of the Cauchy--Green tensor, but rather from this matrix pre-multiplied by the inverse of the coordinate representation of the manifold's metric. This metric is derived via pullback by the coordinate map of the ambient Euclidean 3-space, and analogously, the Cauchy--Green tensor is obtained via pullback by the flow map of this metric. Through these derived eigenvectors, an orthogonality relationship emerges with respect to the manifold's metric. Employing a Riemannian manifold framework required the adaptation of the concept of objectivity to demonstrate that the Cauchy--Green tensor, and hence geodesic vortex detection, is observer-independent. Furthermore, a recently proposed birth-and-death framing algorithm, rooted in geodesic vortex detection, required adaptation to accommodate the restricted temporal applicability of two-dimensional motion in inherently 3-d flows, such as those observed in the Earth's stratosphere, which served as the motivation for our research.

With the aforementioned modifications, our investigation concentrated on the Lagrangian, i.e., kinematic, aspects of the austral stratospheric polar vortex during the extraordinary sudden warming event of 2002. The distinctiveness of this event lies in the splitting of the vortex. To execute this analysis, we implemented geodesic vortex detection on reanalyzed wind data over an isentropic hemispherical cap, centered at the South Pole within the middle stratosphere. A comprehensive analysis of the vortex's life cycle was subsequently presented, encompassing its birth, the splitting process, and its eventual death. We highlight the novel kinematic perspectives provided on ozone depletion within the vortex. Specifically, ozone depletion initiates within a narrow ring located on the poleward side of the vortex. The necessary confinement for ozone-depleting chemical reactions is facilitated by the vortex's edge, identified as a $p$-loop with $p \approx 1$ over a month-long period during which diapycnic mixing is negligible, alongside a $p$-loop with an amplified $p$-value, $p\approx 1.6$. Consequently, ozone-deficient air predominantly mixes poleward, as opposed to equatorward, ultimately leading to the formation of a ``hole'' encompassing the entire breadth of the vortex.

\section*{Acknowledgments}

We extend our gratitude to Daniel Karrasch (Facilitator for Self-Directed Education) for many useful discussions on differential geometry.  We also thank Katherin Padberg-Gehle (Leuphana Universit\"at L\"uneburg) for sharing with us an earlier version (ERA Interim) of the isentropic winds ultimately analyzed in this paper. FJBV recognizes Michael Brown (Rosenstiel School of Marine, Atmospheric \& Earth Science) for influencing his interest in the Lagrangian dynamics of stratospheric polar vortices and ozone holes.

%

\section*{Author Declarations}

\subsection*{Conflict of Interest}

The authors have no conflict of interest to disclose.

\subsection*{Author Contributions}

FAC carried out geodesic vortex detection.  FJBV is responsible for the extension of geodesic vortex detection to curved surfaces. GB performed the adaptation of the \href{https://julialang.org/}{Julia} code of the \href{https://github.com/CoherentStructures/CoherentStructures.jl}{CoherentStructures.jl} package to work on curved surfaces. The manuscript was collaboratively written by FAC, FJBV, and GB.

\appendix
\numberwithin{equation}{section}

\section{Additional details regarding \texorpdfstring{$p$-loops}{p-loops}}\label{app:loops}

A number of observations regarding $p$-loops merit attention.
\begin{itemize}
    \item To show that $r' = \ell_p^\pm(r)$ satisfies $L(r, r') = p$ one must follow \cite{Haller-Beron-13} by proposing $r' = \alpha \nu_{(1)}(r) + \beta \nu_{(2)}(r)$, and subsequently determining the constants $\alpha, \beta$ subject to the constraint $\|r'\|_r = 1$. Keeping the orthonormality relationship \eqref{eq:ortho} in mind, one finds, on one hand $L(r, r')^2 - p^2 = \lambda_1\alpha^2 + \lambda_2\beta^2 - p^2 = 0$ and on the other hand $\|r'\|_r^2 = \alpha^2 + \beta^2 = 1$, just as in \cite{Haller-Beron-13}.  This implies \eqref{eq:p-line}, which is equivalent to the $p$-loop equation(s) that appear in \cite{Haller-Beron-13}. It should be noted that this equivalence is only formal, since $(\lambda_i(r),\nu_i(r))$ is not the $i$th eigenvalue--eigenvector pair of $C(r)$, the coordinate coordinate representation of the Cauchy--Green tensor, as it would be in the $M = \mathbb R^2$ case, but rather of $G^{-1}(r)C(r)$.\footnote{In Appendix C of \cite{Haller-Beron-12}, the extension to Riemannian manifolds of the geodesic theory of transport barriers mentions the matrix relevant for eigenvalue--eigenvector computation. It is noted that this should be $G^{-1}C$ instead of $C$, to align with accurate computation procedures.} 

    \item The limit cycle solutions to \eqref{eq:p-line} either expand or contract with changes in $p$, creating smooth annular regions of non-intersecting loops. This phenomenon occurs because $\ell_p^\pm(x)$ in \eqref{eq:p-line} at each $x \in \mathcal D$ rotates in the same direction with changes in $p$. This allows the application of results from \cite{Duff-53}, at least locally, near a limit cycle where $\ell_p^\pm(x)$ can be smoothly oriented as elaborated in Appendix C of the Online Supplementary Materials of \cite{Haller-Beron-12}.

   \item A necessary condition for the existence of $p$-loops is given by the index theory for planar line fields \cite{Karrasch-etal-14, Karrasch-Schilling-20}.  A subregion $\mathcal D_s \subset \mathcal D$ necessarily supports a $p$-loop if the index or winding number $\text{ind}_{\partial\mathcal D_s} (\ell_p^\pm) = 1$ for either the $+$ or $-$ sign, that is, if $\ell_p^+$ or $\ell_p^-$ makes one turn during one anticlockwise revolution along $\partial\mathcal D_s$. This necessary condition is achieved whenever the number of wedges ($\text{ind} = \frac{1}{2}$) included in $\mathcal D_s$ exceeds that of trisectors ($\text{ind} = -\frac{1}{2}$) by 2.  Wedges and trisectors are the only type of singular points, analogous to critical points of vector fields, in 2-space dimensions (cf.\ \cite{Karrasch-etal-14, Karrasch-Schilling-20} for details). Singular points $x_0^* \in \mathcal D_s$ satisfy 
   \begin{equation}
       C^{t_0,t_1}(x_0^*) = G(x_0^*) \Longleftrightarrow \mathsf C^{t_0,t_1}(x_0^*) = \mathsf I.
   \end{equation} 

   \item A significant observation made in \cite{Haller-Beron-13} is that $p$-loops with $p = 1$, i.e., $1$-loops, demonstrate a complete resistance to the typically observed stretching of arbitrary material loops in turbulent flows. These loops return to their original perimeter at time $t_0$ when observed at time $t_1$. This restoration of the perimeter, alongside the conservation of the enclosed area in the incompressible case, confers exceptional coherence to CLVs encapsulated by $1$-loops. This observation suggests a methodological strategy for identifying CLVs by initiating the exploration of the allowable $p$-range within the proximity of $p = 1$. 

   \item Finally, the boundaries of CLVs in planar flows demonstrate resistance to both stretching and some degree of diffusion \cite{Haller-etal-18}. This trait is also expected for CLVs in flows on curved surfaces. Such a trait can be anticipated in material lines that remain unfolded.
\end{itemize}

\section{Numerical implementation of geodesic vortex detection}\label{app:num}

To facilitate the implementation of geodesic vortex detection in the programming language of the user's choice, we review the basic algorithmic steps involved, adapted for flows on curved surfaces. The \href{https://julialang.org/}{Julia} code of the \href{https://github.com/CoherentStructures/CoherentStructures.jl}{CoherentStructures.jl} package, employed in this paper, was accordingly modified.
\begin{enumerate}
    \item Provide velocity data $u(x,t)$ with $(x,t) \in (\mathcal D,\mathcal I) \subset M \times \mathbb R$ where the coordinate representation of the metric on the Riemannian manifold $M$ is given by $G(x)$.  
    
    \item  Fix $t_0 \in \mathcal I$ and choose some $T$, of absolute value smaller than $|\mathcal I|$.
    
    \item Integrate $u(x,t)$ over $t\in [t_0,t_1]$, $t_1=t_0+T$, for initial conditions $x_0$ on a fine grid $\mathcal G$ covering $\mathcal D$ to evaluate $F^{t_0,t_1}(x_0)$ using some high-order method, such as the Runge--Kutta family.
    
    \item Compute the derivatives of $F^{t_0,t_1}(x_0)$ using finite differences over $\mathcal G$, construct the matrix $G(x_0)^{-1}C^{t_0,t_1}(x_0)$ and compute its eigenvalue--eigenvector pairs ($\lambda_i(x_0),\boldsymbol\nu_{(i)}(x_0))$, $i=1,2$. (Alternatively, one can construct the metric representation of the Cauchy--Green tensor, $\mathsf C^{t_0,t_1}(x_0)$, and compute its eigenvalue--eigenvector pairs ($\lambda_i(x_0),\bbnu_{(i)}(x_0))$.  Then one has the relationship $\boldsymbol\nu_{(i)}(x_0) = \smash{G(x_0)^{-1/2}}\bbnu_{(i)}(x_0)$.  Alternatively, one may choose to utilize $\{\bbnu_{(i)}(x_0)\}$ to directly compute one eigenvector, while determining the other eigenvector through an orthogonal rotation. Should this approach be adopted, the $p$-loops will be represented in the metric space, necessitating a transformation via $G(x_0)^{-1/2}$ to revert to the coordinate space. We have chosen to take this route in writing the \href{https://julialang.org/}{Julia} code \href{https://github.com/70Gage70/CoherentLagrangianVortices}{clv.jl} to take advantage of the existing package \href{https://github.com/CoherentStructures/CoherentStructures.jl}{CoherentStructures.jl}; cf.\ Data Availability section, below.)
    
    \item Construct the coordinate representation of the $p$-line fields $\boldsymbol\ell_p^\pm$, i.e., using $\boldsymbol\nu_{(i)}(x_0)$. This ensures that, when integrated, the $p$-lines lie in coordinate, rather than metric, space.
    
    \item Identify subregions $\mathcal D_s \subset \mathcal D$ where the number of wedge singular points of $G(x_0)^{-1}C^{t_0,t_1}(x_0)$ (or equivalently of $\mathsf C^{t_0,t_1}(x_0)$) exceeds that of the trisector singularities by 2.
   
    \item Identify fixed points in Poincare sections transverse to $\ell_p^\pm$ in each $\mathcal{D}_s$ by varying $p$ and retaining the outermost one. Because of the physical relevance of $p$-loops with $p=1$ the search should begin near $p = 1$ and can progressively expand around the vicinity of $p = 1$ within the allowable $p^2$-range $[\max\lambda_1(r), \min\lambda_2(r)]$ if no cycles are detected. Each corresponding limit cycle defines the boundary of a CLV that resists stretching over $t \in [t_0, t_1]$, with those corresponding to $p \approx 1$ being the most resistant to stretching.
\end{enumerate}

\section*{Data Availability}

The \href{https://julialang.org/}{Julia} package \href{https://github.com/CoherentStructures/CoherentStructures.jl}{CoherentStructures.jl} was created by Daniel Karrasch.  A basic working script with the required adaptations to flows on curved surfaces is distributed via \href{https://github.com/70Gage70/CoherentLagrangianVortices}{https://github.com/70Gage70/CoherentLagrangianVortices}. The wind velocity and ozone concentration originate from the European Centre for Medium-Range Weather Forecasts (ECMWF) Reanalysis v5 (ERA5), which can be accessed via \href{https://www.ecmwf.int/en/forecasts/dataset/ecmwf-reanalysis-v5}{https://www.ecmwf.int/en/forecasts/dataset/ecmwf-reanalysis-v5}.

\bibliographystyle{alpha}

\begin{thebibliography}{ACBVG{\etalchar{+}}22}

\bibitem[AB22]{Andrade-Beron-22}
F.~Andrade‐Canto and F.~J. Beron‐Vera.
\newblock {Do eddies connect the tropical Atlantic Ocean and the Gulf of
  Mexico?}
\newblock {\em Geophysical Research Letters}, 49:e2022GL099637, 2022.

\bibitem[ACBVG{\etalchar{+}}22]{Andrade-etal-22}
Fernando Andrade-Canto, Francisco~J. Beron-Vera, Gustavo~J. Goni, Daniel
  Karrasch, Maria~J. Olascoaga, and Joquin Trinanes.
\newblock {Carriers of \emph{Sargassum} and mechanism for coastal inundation in
  the Caribbean Sea}.
\newblock {\em Phys.\ Fluids}, 34:016602, 2022.

\bibitem[ACKBV20]{Andrade-etal-20}
F.~Andrade-Canto, D.~Karrasch, and F.~J. Beron-Vera.
\newblock {Genesis, evolution, and apocalyse of Loop Current rings}.
\newblock {\em Phys. Fluids}, 32:116603, 2020.

\bibitem[AFK24]{Atnip-etal-24}
Jason Atnip, Gary Froyland, and P\'eter Koltai.
\newblock An inflated dynamic laplacian to track the emergence and
  disappearance of semi-material coherent sets.
\newblock arXiv.2403.10360, 2024.

\bibitem[AHL87]{Andrews-etal-87}
D.~G. Andrews, J.~R. Holton, and C.~B. Leovy.
\newblock {\em {Middle Atmosphere Dynamics}}, volume~40 of {\em International
  Geophysical Series}.
\newblock Academic, 1987.

\bibitem[AKN06]{Arnold-etal-06}
V.~I. Arnold, V.~V. Kozlov, and A.~I. Neishtadt.
\newblock Mathematical aspects of classical and celestial mechanics.
\newblock In {\em Dynamical Systems {III}}, volume~3 of {\em Encyclopedia of
  Mathematical Sciencies}, page 518. Springer-Verlag, Berlin Heidelberg, third
  edition, 2006.

\bibitem[AMR88]{Abraham-Marsden-Ratiu-88}
R.~Abraham, J.~E. Marsden, and T.~Ratiu.
\newblock {\em Manifolds, Tensor Analysis and Applications}.
\newblock Second {E}dition, {A}pplied {M}athematical {S}ciences \textbf{75}.
  Springer, 1988.

\bibitem[Arn73]{Arnold-73}
V.~I. Arnold.
\newblock {\em Ordinary Differential Equations}.
\newblock Massachussets Institute of Technology, 1973.

\bibitem[BSH{\etalchar{+}}15]{Butler-etal-15}
Amy~H. Butler, Dian~J. Seidel, Steven~C. Hardiman, Neal Butchart, Thomas
  Birner, and Aaron Match.
\newblock Defining sudden stratospheric warmings.
\newblock {\em Bulletin of the American Meteorological Society},
  96(11):1913–1928, November 2015.

\bibitem[BSR{\etalchar{+}}24]{Bodnariuk-etal-24}
N.~Bodnariuk, M.~Saraceno, L.~A. Ruiz‐Etcheverry, C.~Simionato,
  F.~Andrade‐Canto, F.~J. Beron‐Vera, and M.~J. Olascoaga.
\newblock Multiple lagrangian jet‐core structures in the {Malvinas Current}.
\newblock {\em Journal of Geophysical Research: Oceans}, 129:e2023JC020446,
  2024.

\bibitem[BVBO{\etalchar{+}}08]{Beron-etal-08-JAS}
F.~J. Beron-Vera, M.~G. Brown, M.~J. Olascoaga, I.~I. Rypina, H.~Kocak, and
  I.~A. Udovydchenkov.
\newblock Zonal jets as transport barriers in planetary atmospheres.
\newblock {\em J. Atmos. Sci.}, 65:3316--3326, 2008.

\bibitem[BVOBK12]{Beron-etal-12}
F.~J. Beron-Vera, M.~J. Olascoaga, M.~G. Brown, and H.~Ko\c{c}ak.
\newblock Zonal jets as meridional transport barriers in the subtropical and
  polar lower stratosphere.
\newblock {\em J. Atmos Sci.}, 69:753--767, 2012.

\bibitem[BVOG10]{Beron-etal-10b}
F.~J. Beron-Vera, M.~J. Olascoaga, and G.~J. Goni.
\newblock Surface ocean mixing inferred from different multisatellite altimetry
  measurements.
\newblock {\em J. Phys. Oceanogr.}, 40:2466--2480, 2010.

\bibitem[BVOH{\etalchar{+}}15]{Beron-etal-15}
F.~J. Beron-Vera, M.~J. Olascoaga, G.~Haller, M.~Farazmand, J.~{Tri\~nanes},
  and Y.~Wang.
\newblock {Dissipative inertial transport patterns near coherent Lagrangian
  eddies in the ocean}.
\newblock {\em Chaos}, 25:087412, 2015.

\bibitem[BVWO{\etalchar{+}}13]{Beron-etal-13}
F.~J. Beron-Vera, Y.~Wang, M.~J. Olascoaga, G.~J. Goni, and G.~Haller.
\newblock {Objective detection of oceanic eddies and the Agulhas leakage}.
\newblock {\em J. Phys. Oceanogr.}, 43:1426--1438, 2013.

\bibitem[CCM21]{Curbelo-etal-21}
Jezabel Curbelo, Gang Chen, and Carlos~Roberto Mechoso.
\newblock Lagrangian analysis of the northern stratospheric polar vortex split
  in april 2020.
\newblock {\em Geophysical Research Letters}, 48(16), August 2021.

\bibitem[CMMW19]{Curbelo-etal-19}
Jezabel Curbelo, Carlos~R. Mechoso, Ana~M. Mancho, and Stephen Wiggins.
\newblock Lagrangian study of the final warming in the southern stratosphere
  during 2002: {Part I. T}he vortex splitting at upper levels.
\newblock {\em Climate Dynamics}, 53:2779–2792, 2019.

\bibitem[dM93]{delCastillo-Morrison-93}
D.~{del-Castillo-Negrete} and P.~J. Morrison.
\newblock Chaotic transport by {R}ossby waves in shear flow.
\newblock {\em Phys. Fluids A}, 5(4):948--965, 1993.

\bibitem[DM08]{Dritschel-McIntyre-08}
D.~G. Dritschel and M.~E. McIntyre.
\newblock {Multiple jets as PV staircases: the Phillips effect and the
  resilience of eddy-transport barriers}.
\newblock {\em J. Atmos. Sci.}, 65:855--874, 2008.

\bibitem[Duf53]{Duff-53}
G.~F.~D. Duff.
\newblock Limit cycles and rotated vector fields.
\newblock {\em Ann. Math.}, 57:15--31, 1953.

\bibitem[EPS06]{Esler-etal-06}
J.~G. Esler, L.~M. Polvani, and R.~K. Scott.
\newblock {The Antarctic stratospheric sudden warming of 2002: A self-tuned
  resonance?}
\newblock {\em Geophys. Res. Lett.}, 33:L12804, 2006.

\bibitem[FBH14]{Farazmand-etal-14}
M.~Farazmand, D.~Blazevski, and G.~Haller.
\newblock Shearless transport barriers in unsteady two-dimensional flows and
  maps.
\newblock {\em Physica D}, 278-279:44--57, 2014.

\bibitem[FGS85]{Farman-etal-85}
J.~C. Farman, B.~G. Gardiner, and J.~D. Shanklin.
\newblock {Large losses of ozone in Antarctica reveal seasonal ClO$_x$/NO$_x$
  interaction}.
\newblock {\em Nature}, 315:207--210, 1985.

\bibitem[FK23]{Froyland-Koltai-23}
Gary Froyland and Péter Koltai.
\newblock Detecting the birth and death of finite‐time coherent sets.
\newblock {\em Communications on Pure and Applied Mathematics},
  76(12):3642–3684, July 2023.

\bibitem[Hal11]{Haller-11}
G.~Haller.
\newblock A variational theory of hyperbolic {Lagrangian Coherent Structures}.
\newblock {\em Physica D}, 240:574--598, 2011.

\bibitem[Hal15]{Haller-15}
G.~Haller.
\newblock Lagrangian coherent structures.
\newblock {\em Ann. Rev. Fluid Mech.}, 47:137--162, 2015.

\bibitem[Hal16]{Haller-16}
G.~Haller.
\newblock Climate, black holes and vorticity: {H}ow on {E}arth are they
  related?
\newblock {\em SIAM News}, 49:1--2, 2016.

\bibitem[Hal23]{Haller-23}
George Haller.
\newblock {\em Transport Barriers and Coherent Structures in Flow Data}.
\newblock Cambridge University Press, Cambridge, United Kindom, 2023.

\bibitem[Hay05]{Haynes-05}
P.~Haynes.
\newblock Stratospheric dynamics.
\newblock {\em Annu. Rev. Fluid Mech.}, 37:263--293, 2005.

\bibitem[HBB{\etalchar{+}}20]{Hersbach-etal-20}
Hans Hersbach, Bill Bell, Paul Berrisford, Shoji Hirahara, András Horányi,
  Joaquín Muñoz‐Sabater, Julien Nicolas, Carole Peubey, Raluca Radu, Dinand
  Schepers, Adrian Simmons, Cornel Soci, Saleh Abdalla, Xavier Abellan,
  Gianpaolo Balsamo, Peter Bechtold, Gionata Biavati, Jean Bidlot, Massimo
  Bonavita, Giovanna Chiara, Per Dahlgren, Dick Dee, Michail Diamantakis,
  Rossana Dragani, Johannes Flemming, Richard Forbes, Manuel Fuentes, Alan
  Geer, Leo Haimberger, Sean Healy, Robin~J. Hogan, Elías Hólm, Marta
  Janisková, Sarah Keeley, Patrick Laloyaux, Philippe Lopez, Cristina Lupu,
  Gabor Radnoti, Patricia Rosnay, Iryna Rozum, Freja Vamborg, Sebastien
  Villaume, and Jean‐Noël Thépaut.
\newblock The {ERA5} global reanalysis.
\newblock {\em Quarterly Journal of the Royal Meteorological Society},
  146:1999--2049, 2020.

\bibitem[HBV12]{Haller-Beron-12}
George Haller and Francisco~J. Beron-Vera.
\newblock Geodesic theory of transport barriers in two-dimensional flows.
\newblock {\em Physica D}, 241:1680--1702, 2012.

\bibitem[HBV13]{Haller-Beron-13}
George Haller and Francisco~J. Beron-Vera.
\newblock {Coherent Lagrangian vortices: The black holes of turbulence}.
\newblock {\em J. Fluid Mech.}, 731:R4, 2013.

\bibitem[HBV14]{Haller-Beron-14}
George Haller and Francisco~J. Beron-Vera.
\newblock {Addendum to `Coherent Lagrangian vortices: The black holes of
  turbulence'}.
\newblock {\em J. Fluid Mech.}, 755:R3, 2014.

\bibitem[HH16]{Hadjighasem-Haller-16}
A.~Hadjighasem and G.~Haller.
\newblock {Geodesic transport barriers in Jupiter's atmosphere: a video-based
  analysis}.
\newblock {\em SIAM Review}, 58:69--89, 2016.

\bibitem[HKK18]{Haller-etal-18}
George Haller, Daniel Karrasch, and Florian Kogelbauer.
\newblock Material barriers to diffusive and stochastic transport.
\newblock {\em Proceedings of the National Academy of Sciences},
  115:9074--9079, 2018.

\bibitem[HY00]{Haller-Yuan-00}
G.~Haller and G.~Yuan.
\newblock Lagrangian coherent structures and mixing in two-dimensional
  turbulence.
\newblock {\em Physica D}, 147:352--370, 2000.

\bibitem[JM87]{Juckes-McIntyre-87}
N.~M. Juckes and M.~E. McIntyre.
\newblock A high-resolution one-layer model of breaking planetary waves in the
  stratosphere.
\newblock {\em Nature}, 328:590--596, 1987.

\bibitem[JRW21]{Jucker-etal-21}
M.~Jucker, T.~Reichler, and D.~W. Waugh.
\newblock How frequent are {Antarctic} sudden stratospheric warmings in present
  and future climate?
\newblock {\em Geophysical Research Letters}, 48(11), June 2021.

\bibitem[Kar15]{Karrasch-15}
D.~Karrasch.
\newblock {Attracting Lagrangian coherent structures on Riemannian manifolds}.
\newblock {\em Chaos}, 25:087411, 2015.

\bibitem[KHH14]{Karrasch-etal-14}
D.~Karrasch, F.~Huhn, and G.~Haller.
\newblock {Automated detection of coherent Lagrangian vortices in
  two-dimensional unsteady flows}.
\newblock {\em Proc. Royal Soc. A}, 471:20140639, 2014.

\bibitem[KS20]{Karrasch-Schilling-20}
Daniel Karrasch and Nathanael Schilling.
\newblock {Fast and robust computation of coherent Lagrangian vortices on very
  large two-dimensional domains}.
\newblock {\em The SMAI Journal of Computational Mathematics}, 6:101--124,
  2020.

\bibitem[LHB{\etalchar{+}}21]{Lim-etal-21}
Eun-Pa Lim, Harry~H. Hendon, Amy~H. Butler, David W.~J. Thompson, Zachary~D.
  Lawrence, Adam~A. Scaife, Theodore~G. Shepherd, Inna Polichtchouk, Hisashi
  Nakamura, Chiaki Kobayashi, Ruth Comer, Lawrence Coy, Andrew Dowdy, Rene~D.
  Garreaud, Paul~A. Newman, and Guomin Wang.
\newblock The 2019 southern hemisphere stratospheric polar vortex weakening and
  its impacts.
\newblock {\em Bulletin of the American Meteorological Society},
  102(6):E1150–E1171, June 2021.

\bibitem[LR10]{Lekien-Ross-10}
F.~Lekien and S.~D. Ross.
\newblock {The computation of finite-time Lyapunoc exponents on unstructured
  meshes and for non-Euclidean manifolds}.
\newblock {\em Chaos}, 20:017505, 2010.

\bibitem[Mat71]{Matsuno-71}
Taroh Matsuno.
\newblock A dynamical model of the stratospheric sudden warming.
\newblock {\em Journal of Atmospheric Sciences}, 28(8):1479 -- 1494, 1971.

\bibitem[McI82]{McIntyre-82}
M.~McIntyre.
\newblock How well do we understand the dynamics of stratospheric warmings?
\newblock {\em J. Meteorol. Soc. Japan}, 60:37--65, 1982.

\bibitem[McI89]{McIntyre-89}
M.~E. McIntyre.
\newblock {On the Antartic ozone hole}.
\newblock {\em J. Atmos. Terr. Phys.}, 51:29--43, 1989.

\bibitem[MR74]{Molina-Rowland-74}
Mario~J. Molina and F.~S. Rowland.
\newblock Stratospheric sink for chlorofluoromethanes: chlorine atom-catalysed
  destruction of ozone.
\newblock {\em Nature}, 249(5460):810–812, June 1974.

\bibitem[MWCM13]{Mancho-etal-13}
Ana~M. Mancho, Stephen Wiggins, Jezabel Curbelo, and Carolina Mendoza.
\newblock Lagrangian descriptors: {A} method for revealing phase space
  structures of general time dependent dynamical systems.
\newblock {\em Comm. Nonlin. Sci. Numer. Sim.}, 18:3530--3557, 2013.

\bibitem[NNRS96]{Nash-etal-96}
Eric~R. Nash, Paul~A. Newman, Joan~E. Rosenfield, and Mark~R. Schoeberl.
\newblock {An objective determination of the polar vortex using Ertel’s
  potential vorticity}.
\newblock {\em Journal of Geophysical Research: Atmospheres},
  101(D5):9471–9478, April 1996.

\bibitem[NPGR21]{Ndour-etal-21}
Moussa Ndour, Kathrin Padberg-Gehle, and Martin Rasmussen.
\newblock Spectral early-warning signals for sudden changes in time-dependent
  flow patterns.
\newblock {\em Fluids}, 6:49, 2021.

\bibitem[NS99]{Ngan-Shepherd-99a}
K.~Ngan and T.~G. Shepherd.
\newblock A closer look at chaotic advection in the stratosphere. {P}art {I}:
  {G}eometric structure.
\newblock {\em J. Atmos. Sci.}, 56:4,134--4,152, 1999.

\bibitem[Ott89]{Ottino-89}
J.~M. Ottino.
\newblock {\em The Kinematics of Mixing: Stretching, Chaos and Transport}.
\newblock Cambridge Texts in Applied Mathematics. Cambridge University Press,
  Cambridge, 1989.

\bibitem[PBB{\etalchar{+}}21]{Polichtchouk-etal-21}
Inna Polichtchouk, Peter Bechtold, Massimo Bonavita, Richard Forbes, Sean
  Healy, Robin Hogan, Patrick Laloyaux, Michael Rennie, Tim Stockdale, Nils
  Wedi, Michail Diamantakis, Johannes Flemming, Stephen English, Lars Isaksen,
  Filip Vána, Sonja Gisinger, and Nicholas Byrne.
\newblock Stratospheric modelling and assimilation.
\newblock ECMWF Technical Memoranda 877, https://www.ecmwf.int/node/19875,
  doi:10.21957/25hegfoq, 2021.

\bibitem[PW04]{Polvani-Waugh-04}
Lorenzo~M. Polvani and Darryn~W. Waugh.
\newblock Upward wave activity flux as a precursor to extreme stratospheric
  events and subsequent anomalous surface weather regimes.
\newblock {\em Journal of Climate}, 17(18):3548 -- 3554, 2004.

\bibitem[RBBV{\etalchar{+}}07a]{Rypina-etal-07a}
I.~I. Rypina, M.~G. Brown, F.~J. Beron-Vera, H.~Kocak, M.~J. Olascoaga, and
  I.~A. Udovydchenkov.
\newblock On the {L}agrangian dynamics of atmospheric zonal jets and the
  permeability of the stratospheric polar vortex.
\newblock {\em J. Atmos. Sci.}, 64:3595--3610, 2007.

\bibitem[RBBV{\etalchar{+}}07b]{Rypina-etal-07b}
I.~I. Rypina, M.~G. Brown, F.~J. Beron-Vera, H.~Ko\c{c}ak, M.~J. Olascoaga, and
  I.~A. Udovydchenkov.
\newblock {Robust transport barriers resulting from strong
  Kolmogorov--Arnold--Moser stability}.
\newblock {\em Phys. Rev. Lett.}, 98:104102, 2007.

\bibitem[SH16]{Serra-Haller-16}
M.~Serra and G.~Haller.
\newblock {Objective Eulerian coherent structures}.
\newblock {\em Chaos}, 26:053110, 2016.

\bibitem[She00]{Shepherd-00}
T.~G. Shepherd.
\newblock The middel atmosphere.
\newblock {\em J. Atmos. Sol.--Terr. Phys.}, 62:1587--1601, 2000.

\bibitem[SKN00]{Shepherd-etal-00}
T.~G. Shepherd, J.~N. Koshyk, and K.~Ngan.
\newblock On the nature of large-scale mixing in the stratosphere and
  mesosphere.
\newblock {\em J. Geophys. Res.}, 105:12433--12466, 2000.

\bibitem[Sol99]{Solomon-99}
S.~Solomon.
\newblock {Stratospheric ozone depletion: A review of concepts and history}.
\newblock {\em Rev. Geophys.}, 37:275--316, 1999.

\bibitem[Spe98]{Speziale-98}
Charles~G. Speziale.
\newblock A review of material frame-indifference in mechanics.
\newblock {\em Applied Mechanics Reviews}, 51:489–504, 1998.

\bibitem[SPHS18]{Shepherd-etal-18}
T.G. Shepherd, I.~Polichtchouk, Robin Hogan, and A.J. Simmons.
\newblock Report on stratosphere task force.
\newblock ECMWF Technical Memoranda 824, https://www.ecmwf.int/node/18259,
  doi:10.21957/0vkp0t1xx, 2018.

\bibitem[SPW05]{Shepherd-etal-05}
G.~Shepherd, T., R.~A. Plumb, and S.~C. Wofsy.
\newblock Preface.
\newblock {\em J. Atmos. Sci.}, 62:565--566, 2005.

\bibitem[SSBVH17]{Serra-etal-17}
M.~Serra, P.~Sathe, F.~J. Beron-Vera, and G.~Haller.
\newblock Uncovering the edge of the polar vortex.
\newblock {\em J. Atoms. Sci.}, 74:3871--3885, 2017.

\bibitem[Ste78]{Stewartson-78}
K.~Stewartson.
\newblock {The evolution of the critical layer of a Rossby wave}.
\newblock {\em Geophys. Astrophys. Fluid Dyn}, 9:185--200, 1978.

\bibitem[Tsi11]{Tsitouras-11}
Ch. Tsitouras.
\newblock {Runge--Kutta pairs of order 5(4) satisfying only the first column
  simplifying assumption}.
\newblock {\em Computers \& Mathematics with Applications}, 62:770–775, 2011.

\bibitem[WOBV15]{Wang-etal-15}
Y.~Wang, M.~J. Olascoaga, and F.~J. Beron-Vera.
\newblock {Coherent water transport across the South Atlantic}.
\newblock {\em Geophys. Res. Lett.}, 42:4072--4079, 2015.

\bibitem[WOBV16]{Wang-etal-16}
Y.~Wang, M.~J. Olascoaga, and F.~J. Beron-Vera.
\newblock {The life cycle of a coherent Lagrangian Agulhas ring}.
\newblock {\em J. Geophys. Res.}, 121:3944--3954, 2016.

\bibitem[WP10]{Waugh-Polvani-10}
D.~W. Waugh and L.~M. Polvani.
\newblock Stratospheric polar vortices.
\newblock In L.~M Polvani, A.~H. Sobel, and D.~W. Waugh, editors, {\em The
  Stratosphere: Dynamics, Chemistry and Transport}, volume 190 of {\em Geophys.
  Monogr.}, pages 43--57, Washington DC, 2010. American Geophysical Union.

\bibitem[WW78]{Warn-Warn-78}
T.~Warn and H.~Warn.
\newblock The evolution of a nonlinear critical level.
\newblock {\em Stud. Appl. Math.}, 59:37--71, 1978.

\end{thebibliography}
\newcommand{\etalchar}[1]{$^{#1}$}

\end{document}